\documentclass[12pt]{article}
\usepackage{xcolor}
\font\elevenbf=cmbx10 scaled\magstephalf 
scaled\magstephalf
\usepackage[centertags]{amsmath}
\usepackage{amssymb}
\usepackage{amsfonts}
\usepackage{epsfig}
\usepackage{amsthm}
\usepackage{cases}
\usepackage{caption}

\usepackage{graphicx}
\usepackage[export]{adjustbox}
\graphicspath{ {./images/} }

\newtheorem{theorem}{{\indent\elevenbf Theorem}}[section]
\newtheorem{lemma}[theorem]{{\indent\elevenbf LEMMA}}
\newtheorem{remark}[theorem]{{\indent\elevenbf Remark}}
\newtheorem{defn}[theorem]{{\indent\elevenbf Definition}}
\newtheorem{coro}[theorem]{{\indent\elevenbf COROLLARY}}
\newtheorem{exam}[theorem]{{\indent\elevenbf Example}}

\newcommand{\bt}{\begin{theorem}}
\newcommand{\et}{\end{theorem}}
\newcommand{\br}{\begin{remark}}
\newcommand{\er}{\end{remark}}
\newcommand{\bd}{\begin{defn}}
\newcommand{\ed}{\end{defn}}
\newcommand{\bl}{\begin{lemma}}
\newcommand{\el}{\end{lemma}}
\newcommand{\bc}{\begin{coro}}
\newcommand{\ec}{\end{coro}}
\newcommand{\bex}{\begin{exam}}
\newcommand{\eex}{\end{exam}}

\newcommand{\be}{\begin{equation}}
\newcommand{\ee}{\end{equation}}
\newcommand{\ba}{\begin{array}}
\newcommand{\ea}{\end{array}}
\catcode`\@=11
\def\theequation{\@arabic{\c@section}.\@arabic{\c@equation}}

\catcode`\@=12 \baselineskip18pt
\title{Equation of State of Hot Neutron Star Matter using Finite Range Simple Effective Interaction }
\author{T. R. Routray$^{1}$*, S. Sahoo$^{1}$, X. Vi\~{n}as$^{2,3,4}$, D. N. Basu$^{5}$, M. Centelles$^{2,3}$\\
$^{1}$School of Physics, Sambalpur University, Burla, Sambalpur - $768019,$ India;\\
$^{2}$Departament de F\'isica Qu\`antica i Astrof\'isica (FQA),\\
 Universitat de Barcelona (UB), Mart\'i i Franqu\`es 1, E-08028 Barcelona, Spain;\\
$^{3}$ Institut de Ci\`encies del Cosmos (ICCUB),\\
Universitat de Barcelona (UB), Mart\'i i Franqu\`es 1, E-08028 Barcelona, Spain;\\
$^{4}$Institut Menorqu\'i d'Estudis, Cam\'i des Castell 28, 07702 Ma\'o, Spain;\\
$^{5}$Variable Energy Cyclotron Centre, 1/AF Bidhan Nagar, Kolkata- $700064,$ India.}
\begin{document}
\renewcommand{\thefootnote}{\fnsymbol{footnote}}
\footnotetext[1]
{E-mail:
trr1@rediffmail.com(corresponding author)}

\date{\today}

\maketitle
\begin{abstract}
{The equation of state of hot neutron star matter of n+p+e+$\mu$ composition in $\beta$-equilibrium is studied for both neutrino-free isothermal and neurino-trapped isoentropic conditions, using the formalism where the thermal evolution is built upon its zero-temperature predictions in a self-consistent manner. The accuracy of the parabolic approximation, often used in the finite temperature calculation of hot neutron star matter, is verified by comparing with the results obtained from the exact evaluation in the neutrino-free neutron star matter.
The equation of state of neutrino-trapped isoentropic matter at low entropic condition, relevant to the core-collapsing supernovae, is formulated. In the isoentropic matter, the particle fractions and equation of state have marginal variance as entropy per particle vary between 1 to 3 (in the unit of k$_B$), but the temperature profile shows marked variation. The isentropes are found to be much less sensitive to the nuclear matter incompressibility, but have large dependence on the slope parameter L. The bulk properties of the neutron stars predicted by the isoentropic equation of states for different entropy are calculated. A model calculation for the early stage evolution of protoneutron star to neutron star configuration is also given.}
\end{abstract}



\bigskip

\section{Introduction}

In the model calculations of the high energy astrophysical events of the binary neutron star merger (BNSM), supernovae explosion and protoneutron
stars (PNSs) the important inputs are the pressure ($P$), energy density ($H$), density ($\rho$) and temperature ($T$). Hence, computation of
equation of state (EoS) of hot neutron star matter (NSM) using microscopic methods
\cite{Bombaci1994,Baldo1999,Zuo2004,Burgio2011,Wellenhofer2015,Drischler2016,Togashi2017,Carbone2018,Carbone2019} as well as effective models
\cite{Lattimer1991,Schneider2017,Lattimer2016,Sugahara1994,Steiner2013,Typel2010,Fattoyev2010} has gained momentum in the recent years. But the
composition of the neutron stars (NSs) is still not understood to a satisfactory level. The EoS of hot NSM is known still to a lesser extent. The
temperature in the event of BNSM producing high oscillation in the merged matter rises above 10 MeV \cite{Baiotti2017,Paschalidis2017}. In PNSs
produced in supernovae explosion, $T$ is above 20 MeV \cite{Lattimer1986,Burrows1988} and these are some
events whose observables could serve as potential constraints for understanding the EoS of hot NSM. But the observables are scanty
and not free from large uncertainties for drawing any decisive conclusion on the EoS of hot NSM. The gravitational waves (GWs) detection from two
NSs merger \cite{Abbott2017,Abbott2020} could serve as potential probes to constrain the EoS of hot and dense NSM. However, the
present limitation of LIGO-Virgo detections, whose highest sensitivity is maximum for frequency $f\leq$ 1 kHz, provide GWs
information mostly concentrated in the inspiraling stage during which the two NSs are expected to be in zero $T$ thermodynamical state.
Under these circumstances, a probable way to proceed is to formulate any possible event that can take place in such environment by
constructing the EoS of hot NSM from reasonable consideration of its composition.
 In this work we have used the widely employed concept of n+p+e+$\mu$ matter. However, we are aware that works taking hyperons and
quark matter have been also reported in earlier literature
\cite{Shibata2005,Bauswein2012,Hotokezaka2013,Bernuzzi2015,Rezzolla2016,Zappa2018,Bauswein2019,Most2020}. It is worth mentioning that construction
of EoS of hot NSM for astrophysical utility was initiated more than half a century ago with the pioneering works
of Bethe et al \cite{Bethe1979}, Brown et al \cite{Brown1982}, Lamb et al \cite{Lamb1978}, Lattimer and Ravenhall \cite{Lattimer1978} and
Lattimer \cite{Lattimer1981} and it has gained momentum over the years. In the domain of non-relativistic (NR)
effective models, EOSs of hot NSM based on the Compressible Liquid Drop Model, formulated by Lattimer and Swesty \cite{Lattimer1991}, have
 been widely used. In the relativistic domain, the effective model of Shen et al \cite{Shen1998} has been applied in studies of BNSM, PNS and
black-hole formation \cite{Shen2011}. The temperature effects on the EoS of NSM has also been investigated in the chiral effective
 field theory \cite{Sammarruca2020}.
The CompOSE online directory for neutron star EoS tables has provided a platform to easily publish finite temperature
EoS tables, and for a
recent review of finite-temperature EoSs, see Ref.\cite{Oertel2017}. Despite these efforts, as pointed out by Raithel et al. \cite{Raithel2021}, the number of
finite-temperature EoS models remains relatively
small and they do not span the full range of possible dense matter
physics. In addition, some models are not consistent
with modern astrophysical constraints. For example, several
of the finite-temperature EoS tables predict cold neutron star
radii of $\geq$13 km (e.g., the NL3, TM1, DD2, and TMA EoSs;
see, e.g., Table 1 of \cite{Fischer2014} and references therein), which are in
tension with the latest constraints inferred from LMXB
observations and from GW170817 \cite{Baiotti2019,Raithel2019,Ozel2016}.

 There is another popular approach, the so-called “hybrid” approach, which was  introduced in Ref.\cite{Janka1993} and is now widely used, where a thermal correction is added to the zero-temperature EoS to account for the temperature effect. In these models the temperature effect is not simulated by the
interaction which is used to study the zero-temperature thermodynamical properties of the system, and hence lacks in consistency between zero- and finite $T$ predictions. The laws of thermodynamics will hold consistently where the hot NM EoS evolves on its zero-temperature counterpart. In this context, Behera and collaborators studied the thermal evolution of the effective mass in symmetric nuclear matter (SNM) using the finite range simple effective interaction (SEI) \cite{Behera2002}.
Later on the study was extended for finite $T$ calculation of symmetry energy \cite{Behera2009} and thermal evolution of nuclear properties of asymmetric nuclear matter (ANM) \cite{Behera2011} under the parabolic approximation (PA). The PA rests on the quadratic approximation of the Taylor expansion of energy in ANM in even-powers of isospin asymmetry, where the symmetry energy is approximated as the difference between the energy per particle in the pure neutron matter (PNM) and the SNM. Thus, under the PA, the study of ANM amounts to the independent studies of SNM and PNM. At zero-temperature, the PA is a powerful approximation giving similar results as that of the exact calculation in ANM as has been verified in the earlier works \cite{Bombaci1991,Prakash1997,Li2009}. But for the finite $T$ calculation the validity of the PA requires an explicit verification, which to our knowledge has not been done although most of the model, both NR as well as relativistic, have used the PA \cite{Nicotra2006,Burgio2007,Burgio2010,Xu2007,Moustakidis2009,Wellenhofer2016,Moustakidis2021,Bombaci2021}.

In the present work, we will evaluate
the beta stability and charge neutrality conditions for hot n+p+e+$\mu$ NSM adopting a self-consistent exact evaluation of the EoS of asymmetric nuclear matter at finite $T$. This study is performed using the SEI model with a Yukawa form-factor, refer to as SEI-Y, that has been employed in earlier studies of, both, cold and hot
NSM and neutron star phenomenology \cite{Behera2002,Behera2009,Behera2011,Trr2023,Trr2024}.
The thermodynamics of NSM under the neutrino- trapped conditions will be worked out and compared with the neutrino-free counterparts. The EoS of
neutrino-trapped isoentropic NSM, relevant to the core-collapse supernovae and protoneutron star, is worked out. Evolution of protoneutron star
to neutron star in the early stage is discussed.
In Section 2 we present the basic formulation and the self-consistent exact evaluation procedure adopted to evaluate the composition and EoS of
charge neutral beta-stable n+p+e+$\mu$ NSM, both, for neutrino-free and trapped conditions. In Section 3, we present our results on the composition
 and thermodynamical properties of neutrino-free isothermal NSM under the exact evaluation of hot ANM.
 The validity of the PA is examined by comparing to the exact results.
The neutrino-trapped isoentropic EoS applicable to the PNS is computed and the early stage evolution PNS to NS is discussed. A brief summary of
the work and future perspective is given in the last Section 4.
\section{Basic Formalism}

 Several studies of astrophysical interest have been performed at NR mean field level using the finite range SEI
\cite{Trr2024,Behera1998,Trr2013}, which is defined as
\begin{eqnarray}
V_\textrm{eff}=
& & t_{0}(1+x_{0}P_{\sigma})\delta(\vec{r}) +\frac{t_{3}}{6}(1+x_{3}P_{\sigma})\left(\frac{\rho(\vec{R})}
{1+b\rho(\vec{R})}\right)^{\gamma}\delta(\vec{r})\nonumber \\
& & + (W+BP_{\sigma}-HP_{\tau}-MP_{\sigma}P_{\tau})f(r)\nonumber\\
& & + \textrm{Spin-orbit part},
\label{SEI}
\end{eqnarray}
where a zero-range spin-orbit (SO) interaction depending on a strength parameter $W_0$ is taken to deal with finite nuclei and $f(r)$ is the finite range form factor that can be either Yukawa/Gaussian/exponential one. Here we have taken the Yukawa form factor, $f(r)=e^{-r/\alpha}/{(r/\alpha)}$, where $\alpha$ is the range parameter. In nuclear matter (NM), SEI has eleven parameters, namely, $\alpha$, $\gamma$, b, $x_{0}$, $t_{0}$, $x_{3}$,
$t_{3}$, W, B, H, and M (an additional SO parameter $W_{0}$ enters when finite nucleus is considered). Nine of these parameters, namely $\gamma$, $b$, $\alpha$, $\varepsilon_{0}^{l}$, $\varepsilon_{0}^{ul}$, $\varepsilon_{\gamma}^{l}$, $\varepsilon_{\gamma}^{ul}$, $\varepsilon_{ex}^{l}$, $\varepsilon_{ex}^{ul}$, formed out of the combination of these eleven parameters are needed in the study of isospin asymmetric nuclear matter. The relation of these new parameters with the interaction parameters are given in \cite{Trr2024}. The parameter fitting protocol, which is same for any of the three functional forms of $f(r)$, is somewhat different from those adopted in other conventional Gogny, Skyrme and M3Y effective interactions and is discussed in detail in Refs.\cite{Trr2013,Trr2015}. In the astrophysical domain SEI  has been used in the studies of mass-radius relation \cite{Behera2007}, thermal evolution of symmetry energy and EoS of hot NSM \cite{Behera2009,Behera2011}, crust-core transition in NS \cite{Mario2019}, $r$-mode oscillations in NS pulsars \cite{Trr2018} and spin-down mechanism in newborn NSs \cite{Trr2021}. In these  works the ability of SEI to predict the momentum dependence of the mean field and density dependence of the EoS in good agreement with the results provided by realistic and microscopic calculations is also shown \cite{Sammarruca2010,Wiringa1988,APR1998}.

The energy density $H_T(\rho_n,\rho_p)$ of ANM at temperature $T$ resulting from the SEI in Eq.(\ref{SEI}) is given by,
\begin{eqnarray}
H_T(\rho_n, \rho_p) &=& \frac{\hslash^2}{2m}\int\left[ f_{T}^n({\bf k})
+f_{T}^p({\bf k})\right]k^2d^3k \nonumber \\
&& + \frac{1}{2}\left[\frac{\varepsilon_0^l}{\rho_0}+\frac{\varepsilon_{\gamma}^l}
{\rho_0^{\gamma+1}}\left(\frac{\rho}{1+b\rho}\right)^{\gamma}\right]
\left(\rho_n^2+\rho_p^2\right) \nonumber \\
&&+\left[\frac{\varepsilon_0^{ul}}{\rho_0}+\frac{\varepsilon_{\gamma}^{ul}}
{\rho_0^{\gamma+1}}\left(\frac{\rho}{1+b\rho}\right)^{\gamma}\right]\rho_n\rho_p \nonumber \\
&& +\frac{\varepsilon_{ex}^l}{2\rho_0}\int\int\left[f_T^n({\bf k})f_{T}^n
({\bf k'})+f_T^p({\bf k})f_{T}^p({\bf k'})\right]
g_{ex}\left(\vert{\bf k}-{\bf k'}\vert\right)d^3k d^3k' \nonumber \\
&& +\frac{\varepsilon_{ex}^{ul}}{2\rho_0}\int\int\bigg[f_T^n({\bf k})f_{T}^p
({\bf k'})+f_T^p({\bf k})f_{T}^n({\bf k'})\bigg]
g_{ex}\left(\vert{\bf k}-{\bf k'}\vert\right)d^3k d^3k',
\label{eq21a}
\end{eqnarray}
 where $\rho_0$ is the saturation density, $f_{T}^{n(p)}$ is the Fermi-Dirac (FD) distribution functions of neutron (proton), $g(|\bf {k}-\bf {k'}|)$ is the normalized Fourier transform of the finite range form factor $f(r)$, which for the Yukawa type is given by
\begin{eqnarray}
g(|\bf {k}-\bf {k'}|)=\frac{1}{1+(\frac{\bf {k}-\bf {k'}}{\Lambda})^2}.
\label{eq3}
\end{eqnarray}
The thermal evolution of the EoS is simulated through the FD distribution functions appearing in the kinetic energy (KE) part and the finite range exchange part of the interaction energy as can be seen from Eq.(\ref{eq21a}). The n (p) FD function is given by,
\begin{eqnarray}
f^{n(p)}_T({k})=\frac{1}{1+exp\left[{\left\{
\epsilon_{T}^{n(p)}(k,\rho_n,\rho_p)-\mu_{T}^{n(p)}
\right\}/T}\right]},
\label{eq6}
\end{eqnarray}
where $\epsilon_T^{n(p)}(k,\rho_n,\rho_p)$ and $\mu_T^{n(p)}$ are the n (p) single particle energy and the chemical potential, respectively, at temperature $T$.
The neutron (proton) single particle energy is obtained by taking the functional derivative of
$H_T$ in (\ref{eq21a}), i.e., $\frac{\partial H_T}{\partial [f_T^{n(p)}]}$.
The neutron single particle energy $\epsilon_T^{n}( k, \rho_n,\rho_p )$ for the energy density in Eq.\ref{eq21a} of SEI is given by,
\begin{eqnarray}
\epsilon_T^{n} ( k, \rho_n,\rho_p )
& &=\frac{\hbar^2 k^2}{2m}
+\varepsilon_{0}^{l} \frac{\rho_n}{\rho_0}
+\varepsilon_{0}^{ul} \frac{\rho_p}{\rho_0}
+\left(\varepsilon_{\gamma}^{l} \frac{\rho_n}{\rho_0}
+\varepsilon_{\gamma}^{ul} \frac{\rho_p}{\rho_0}\right)\left(\frac{\rho}{(1+b\rho)}\right)^{\gamma}\nonumber \\
& &+\frac{\varepsilon_{ex}^{l}} {\rho_0}\int{f_{T}^{n} (\bf {k'}) g(|\bf {k}-\bf {k'}|} d^3 {\bf {k'}}\nonumber \\
& &+\frac{\varepsilon_{ex}^{ul}} {\rho_0}\int{f_{T}^{p} (\bf {k'}) g(|\bf {k}-\bf {k'}|} d^3 {\bf {k'}}\nonumber \\
& &+\left[ \frac{\varepsilon_{\gamma}^{l}}{2\rho_0^{\gamma+1}}\left(\rho_n^2+\rho_p^2\right) +
\frac{\varepsilon_{\gamma}^{ul}}{\rho_0^{\gamma+1}}\left(\rho_n^2 \rho_p^2\right)
 \right] \frac{\gamma \rho^{\gamma-1} }{\left(1+b\rho \right)^{\gamma+1}}
\label{eq4}
\end{eqnarray}
where, the last term is the single particle rearrangement energy. The proton single particle energy can be written from Eq.(\ref{eq4}) by interchanging n and p.
The n and p FD functions are subject to the normalizations,
\begin{eqnarray}
\rho_n=\frac{\xi}{(2\pi)^3}\int{f_T^n(k) d^3k},   {\text {  and  }}   \rho_p=\frac{\xi}{(2\pi)^3}\int{f_T^p(k) d^3k}.
\label{eq5}
\end{eqnarray}
where $\xi$=2 is the spin degeneracy factor.
The evaluation of FD function $f_T^{n(p)}(k)$ requires the knowledge of $\epsilon_{T}^{n(p)}$ which, in turn, involves the distribution function, and therefore requires a self-consistent calculation. To do that we proceed as follows. For given values of the density $\rho$, isospin asymmetry $\beta$=$(\rho_n-\rho_p)/(\rho_n+\rho_p)$ and temperature $T$, the zero-temperature $\epsilon_{0}^{n(p)}(k,\rho_n,\rho_p)$ is used as input in Eq.(\ref{eq6})
to obtain the $\mu_{T}^{{n}^{(0)}}$ and $\mu_{T}^{{p}^{(0)}}$ by solving the two equations in (\ref{eq5}).
 With these values, the zeroth-order n,p-FD
distribution functions $f^{{n}^{(0)}}$, $f^{{p}^{(0)}}$ are obtained from Eq.(\ref{eq6}), which in turn are used to evaluate the
first order single particle energies and the chemical potentials
with the help of Eqs.(\ref{eq4}) and (\ref{eq5}), respectively.
The iteration continues till convergence is reached for which we have used the condition that the chemical potential differences in both $\mu_{T}^{n}$ and $\mu_{T}^{p}$ between two consecutive iterations become less than 0.01 percent. This procedure of self-consistent evaluation of the n (p) FD functions $f^{n(p)}_T({k})$ at finite $T$ provides simultaneously, the n (p) chemical potential $\mu_{T}^{n(p)}$ and single particle energy $\epsilon_{T}^{n(p)}$, and enables to compute the thermodynamical quantities of hot ANM.

\subsection{Hot Neutron star matter EOS}

The outer core of the neutron star comprises of mostly neutrons in fluid state with a few percentage of protons in charge neutral $\beta$-stable condition. The presence of hyperons and exotic matter in the inner core is still in a stage of exploration by considering these possibilities connecting to the well-accepted observables \cite{Shibata2005,Rezzolla2016,Zappa2018,Bauswein2019,Most2020}.
In this work we shall restrict our study to the $n+p+e+\mu$ composition of NSM. The $\beta$-equilibrium is established between the weak decay processes of neutron and the lepton capture of proton,
\begin{eqnarray}
n \rightarrow p+l^{-}+ \bar{\nu_{e}},  {  \text { and }  }
 p+l^{-} \rightarrow n+{\nu_{e}}
\label{eq7}
\end{eqnarray}
In order to remain in $\beta$-equilibrium condition, the chemical potentials of the particles need to obey the relation
\begin{eqnarray}
\mu_{n} = \mu_{p}+\mu_{e}+\mu_{\bar{\nu_{l}}},   {  \text { and }  }
 \mu_{p} + \mu_{e}=\mu_{n}+\mu_{\nu_{l}},
\label{eq8}
\end{eqnarray}
where $\mu_{l}$ and $\mu_{\nu_{l}}$ ($\mu_{\bar{\nu_{l}}}$) are the chemical potentials of lepton and lepton neutrino (anti-neutrino).
Now, there are two situations that one may encounter, (a) 'neutrino-free' condition where the medium is transparent to the neutrinos produced in the reactions, a situation encountered in the cold NSs and (b) 'neutrino-trapped' case where the emitted neutrinos are trapped in the system, a situation that occurs, for example, in the core-collapse supernovae and protoneutron stars or in the two NSs merger remnants.

In the neutrino free case, as the neutrinos escape the system, they do not contribute to the thermodynamics as well as to the EOS of the system. Here the $\beta$-equilibrium condition is
\begin{eqnarray}
{\mu_{n}}-{\mu_{p}} = {\mu_{e}} = {\mu_{\mu}}, { \text {  } }
Y_{p} =Y_{e}+Y_{\mu},
\label{eq91}
\end{eqnarray}
where the second equation is for the charge neutrality with
$Y_{i}$ = $\rho_{i}/\rho$, $i$= p,e and $\mu$, are the proton, electron and muon fractions. The muons are produced when the chemical potential difference $(\mu_{n}-\mu_{p})$ exceeds the rest mass energy of muon, i.e. $(\mu_{n}-\mu_{p}) \ge m_{\mu} c^{2}$ = 105.658 MeV. The exact way of solving the $\beta$-equilibrium charge neutrality conditions in Eq.(\ref{eq91}) at finite $T$ is by self-consistently evaluating $\mu_n$ and $\mu_p$ using the EoS of ANM, given in the forgoing discussion, on varying $Y_p$ till the two conditions in Eq.(\ref{eq91}) are simultaneously satisfied.
A popular way of presenting the $\beta$-equilibrium condition is to use the PA for the nuclear EoS. Under PA at finite $T$, $({\mu_{n}}-{\mu_{p}})$ can be expressed in terms of nuclear free symmetry energy $F_{sym}(\rho)$, i.e.,\cite{Nicotra2006,Burgio2007,Moustakidis2021}
\begin{eqnarray}
 {\mu_{n}}-{\mu_{p}} = 4(1-2Y_{p}) F_{sym}(\rho),
	\label{eq9}
\end{eqnarray}
	In the PA, the free symmetry energy is approximated by the difference of free energy per particle $\bar{F}^N$ in pure neutron matter and $\bar{F}$, that of the symmetric nuclear matter,
	\begin{eqnarray}
	F_{sym}(\rho)=\bar{F}^{N}(\rho)- \bar{F}(\rho),
		\label{eq10}
\end{eqnarray}
which in the T=0 reduces to the symmetry energy. As mentioned before, many works in NSM  at finite $T$ use this PA \cite{Nicotra2006,Burgio2007,Xu2007,Moustakidis2009,Moustakidis2021,Bombaci2021}.
	
Once the charge neutrality and $\beta$-equilibrium conditions in Eq.(\ref{eq91}) are solved for a given density $\rho$ and temperature $T$, the different equilibrium particle fractions $Y_{i}$, $i=n,p,e,{\mu}$, are known and then the EoS of NSM at temperature $T$ can be obtained from the relations,
		\begin{eqnarray}
	F_{NSM}=F_{N}(\rho,Y_{p},T)+F_{e}(\rho_{e},Y_{e},T)+F_{\mu}(\rho_{\mu},Y_{\mu},T)\nonumber\\
	P_{NSM}=P_{N}(\rho,Y_{p},T)+P_{e}(\rho_{e},Y_{e},T)+P_{\mu}(\rho_{\mu},Y_{\mu},T),
		\label{eq11}
\end{eqnarray}
where, F and P, with the sub-script N, e and $\mu$, are the free energy density and pressure denoting those of nucleonic part and leptonic part of both the species e and $\mu$. The leptonic system of e and $\mu$ are treated under the relativistic non-interacting Fermi gas model.
In Eq.(\ref{eq11}) the nucleonic free energy density $F_{N}$ is given by:
		\begin{eqnarray}
	F_{N}(\rho,Y_{p},T) = H_{N}(\rho,Y_{p},T) - T(S_{n} +S_{p}),
			\label{eq12}
\end{eqnarray}
	where, $H_{N}$ is the nucleonic energy density of ANM at temperature $T$ in Eq.(\ref{eq21a}), computed at the equilibrium $Y_p$-value. In this Eq.(\ref{eq12}) $S_n$ ($S_p$) is the neutron (proton) entropy density defined as
	
\begin{eqnarray}
S_{n(p)}=-\frac{\xi}{(2\pi)^3}\int{[f_T^{n(p)}(k) ln f_T^{n(p)}(k) + (1-f_T^{n(p)}(k)) ln(1- f_T^{n(p)}(k))]d^3k},
\label{eq13}
\end{eqnarray}
with $\xi$=2 for n, p, e and $\mu$, whereas, it is 1 for neutrinos. Finally, the nucleonic contribution to the EoS of NSM, i.e. the nuclear pressure, reads:
\begin{eqnarray}
P_N(\rho,Y_p)=\mu_n\rho_n+\mu_p\rho_p - F_N(\rho,Y_p).
\label{eq14}
\end{eqnarray}

The leptonic contribution to the EoS on NSM is built up in a similar way. The energy density for each kind of leptons is provided
by the relativistic non-interacting Fermi-gas model, which reads
  \begin{eqnarray}
  H_{l} =\int{\sqrt{c^{2}\hbar^{2}k^{2}+m_{l}^{2}c^{4}}{\text{ }} f_{T}^{l}(k) d^{3}k},
	\label{eeq15}
  \end{eqnarray}
	where, $f_{T}^{l}(k)=1/ [{1+exp\{[\sqrt{c^{2}\hbar^{2}k^{2}+m_{l}^{2}c^{4}}-\mu_{T}^{l}]/T\}}]$
  is the leptonic distribution function with $l=e  {  \text {   or  }  }  \mu$.
  The EoS provided by each type of leptons is given by
  \begin{equation}
  P_{l} = \mu_{l}\rho_{l}-H_{l}+TS_{l},
	\label{eq15}
  \end{equation}
  where the lepton entropy density $S_{l}$, $l$=$e,\mu$, is obtained using Eq.(\ref{eq13}) with the nucleon distribution functions
$f_T^{n(p)}$ replaced by the leptonic ones $f_{T}^{l}$.
	
   The neutrino-trapped condition occurs in supernovae matter, protoneutron stars and in merger of two NSs where under the high temperature condition the compressed matter becomes opaque to the neutrinos. Due to the trapped neutrinos, which now contribute to the EoS of NSM, the $\beta$-equilibrium condition in Eq.(\ref{eq8}) becomes,
  \begin{equation}
  \mu_{n}-\mu_{p} = \mu_{e} - \mu_{\nu_{e}} = \mu_{\mu} - \mu_{\nu_{\mu}},
	\label{eq16}
  \end{equation}
where the neutrino and antineutrino chemical potentials are related by $\mu_{\nu}=-\mu_{\bar{\nu}}$,
and the charge neutrality condition remains the same $Y_{p}=Y_{e}+Y_{\mu}$.
 The EoS of NSM will now have additional contributions from the lepton
neutrinos and antineutrinos, which read
\begin{equation}
 F_{\nu_{l}}=H_{\nu_{l}}-TS_{\nu_{l}}, \quad  {\text {and}} \quad  P_{\nu_{l}}=\mu_{\nu_{l}}\rho_{\nu_{l}}-F_{\nu_{l}},
\label{eq16a}
\end{equation}
which have to be added to equations (\ref{eq11}).

\section{Results and discussions}

 The thermodynamics of $\beta$-equilibrated NSM is computed using the EoS of SEI-Y for which $e(\rho_0)$= -16 MeV, $\gamma$ =2/3 ($K(\rho_0)\simeq$ 254 MeV),
 ${T}_{f0}$=37 MeV ($\rho_{0}$=0.16103 $fm^{-3}$) and $E_{s}(\rho_{0})$= 30 MeV. The parameters of SEI and the other predicted saturation properties are given in Table-\ref{tab1}.
\begin{table}[h]
	\caption{\small{Nine numbers of parameters for ANM of SEI-Y interaction set alongwith the nuclear matter saturation properties }}
     \centering
	\begin{scriptsize}
	\begin{tabular}{cccccccccc}
		\hline\hline
		$Interaction$&$\gamma$&b[$fm^{3}$]&$\alpha$[fm]&$\varepsilon_{ex}$[MeV]
        &$\varepsilon_{ex}^{l}$[MeV]&$\varepsilon_{0}$[MeV]&$\varepsilon_{0}^{l}$[MeV]&$\varepsilon_{\gamma}$[MeV]&$\varepsilon_{\gamma}^{l}$[MeV]\\

		\hline
		$SEI-Y$&2./3.& 0.7607&0.4232&-129.25&-86.165&-33.676&-42.895&57.934&61.749\\
		\hline
		\hline
		\hline
		\multicolumn{8}{c}{\textbf{Nuclear matter saturation properties}}\\
		\hline
	$Interaction$&	$\rho_{0} [fm^{-3}]$&e$(\rho_{0})$[MeV]&K[MeV]&m*/m&E$_{s}$[MeV]&L[MeV]&K$_{sym}$&Q$_{sym}$\\
		\hline
		$SEI-Y$&0.16103&-16.0&253.6&0.686&30&75.0&-37.159&119.15\\
		\hline\hline
		\label{tab1}	
	\end{tabular}
     \end{scriptsize}
     \end{table}
		The symmetry incompressibility, $K_{sym}$, and skewness, $Q_{sym}$, reported in Table \ref{tab1} are defined as
		\begin{eqnarray}
		K_{sym}=9\rho_{0}^{2}\frac{\partial^{2}E_{sym}(\rho)}{\partial\rho^{2}}|_{\rho=\rho_{0}},
        \text{ and }
       Q_{sym}=27\rho_{0}^{3}\frac{\partial^{3}E_{sym}(\rho)}{\partial\rho^{3}}|_{\rho=\rho_{0}}\nonumber
        	\end{eqnarray}
		respectively.
\subsection{Neutrino-free hot NSM}

\begin{figure}[t]
	\begin{center}
		\includegraphics[height=11cm,width=18cm]{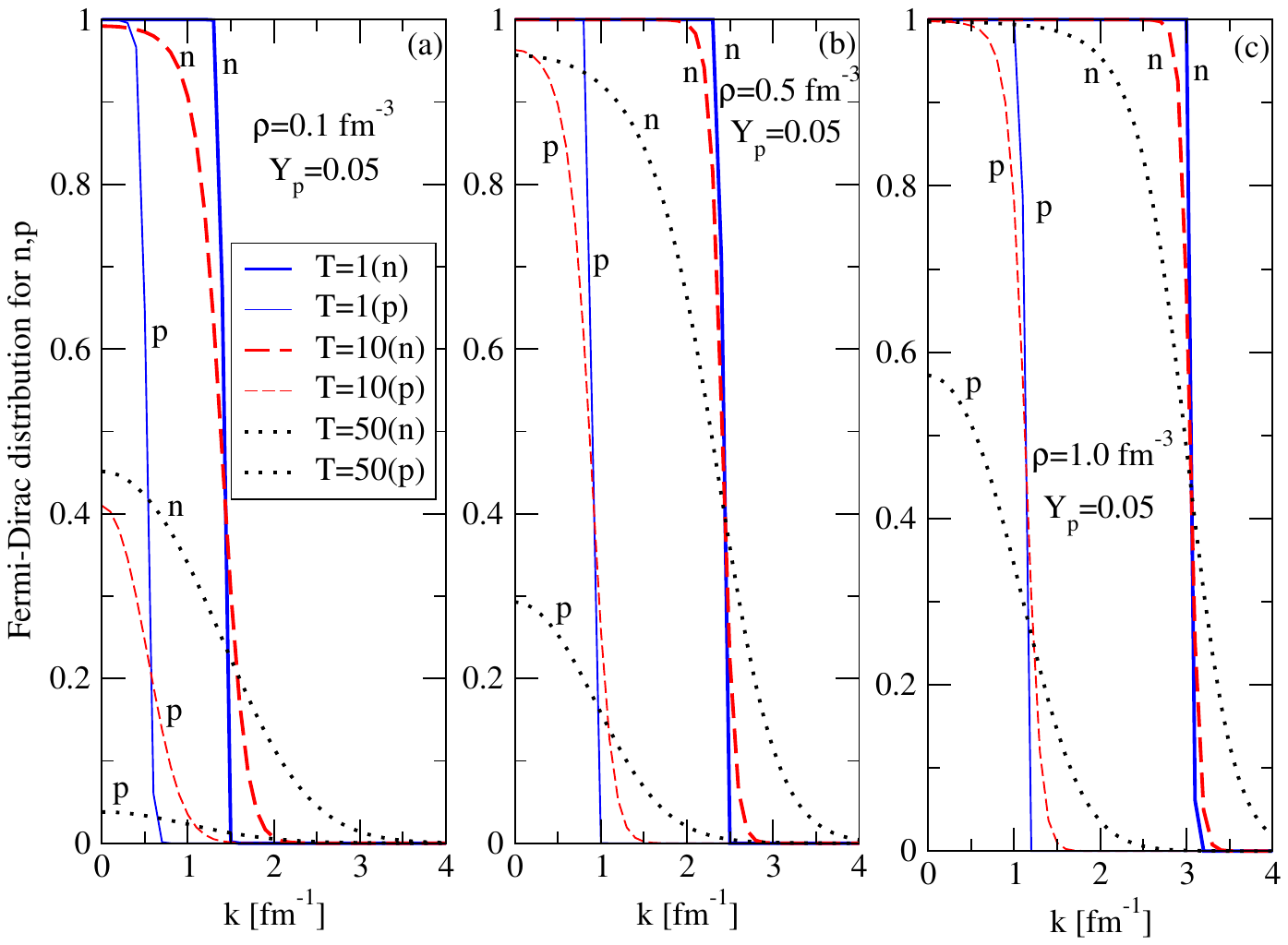}

		\caption{FD distributions for neutron and proton at $T$=1,10 and 50 MeV as a function of momentum $k$ for densities $\rho$=0.1, 0.5 and 1.0 fm$^{-3}$ and isospin asymmetry $\beta$=0.05, shown in three panels, obtained from the exact evaluation of ANM at finite $T$. Legends used are the same in the three panels.}
		\label{fig1}
	\end{center}
\end{figure}

 In this subsection, we shall consider the neutrino-free $\beta$-equilibrated hot NSM and thereafter the neutrino-trapped case. The study is made by evaluating the finite temperature EoS of ANM exactly. Side-by-side evaluation of the EoSs of PNM and SNM separately is made as required under the parabolic approximation. In order to visualize the effect of temperature on the n,p-systems in ANM, the FD distribution functions resulting from the exact evaluation, are computed with a proton fraction $Y_{p}$=0.05 are shown in Figure \ref{fig1} as function of momentum $k$ at temperature $T$=1, 10 and 50 MeV for the total densities $\rho$ = 0.1,0.5 and 1.0 fm$^{-3}$. At $T$=1 MeV, the neutron FD distributions for all three densities are perfectly step-functions. At this temperature $T$=1 MeV, the proton FD distribution of $\rho$=0.1fm$^{-3}$ is depleted at Fermi surface to a small extent because in this case its Fermi energy corresponding to the $T$=0 distribution becomes comparable to 1 MeV.
With growing temperature, the n,p Fermi surfaces are depleted to different extents because nucleons are occupying higher excited states depending on the density $\rho$ and temperature $T$, as can be seen in Fig.\ref{fig1}.
%
%

\begin{figure}[t]
	\begin{center}
		\includegraphics[height=11cm,width=18cm]{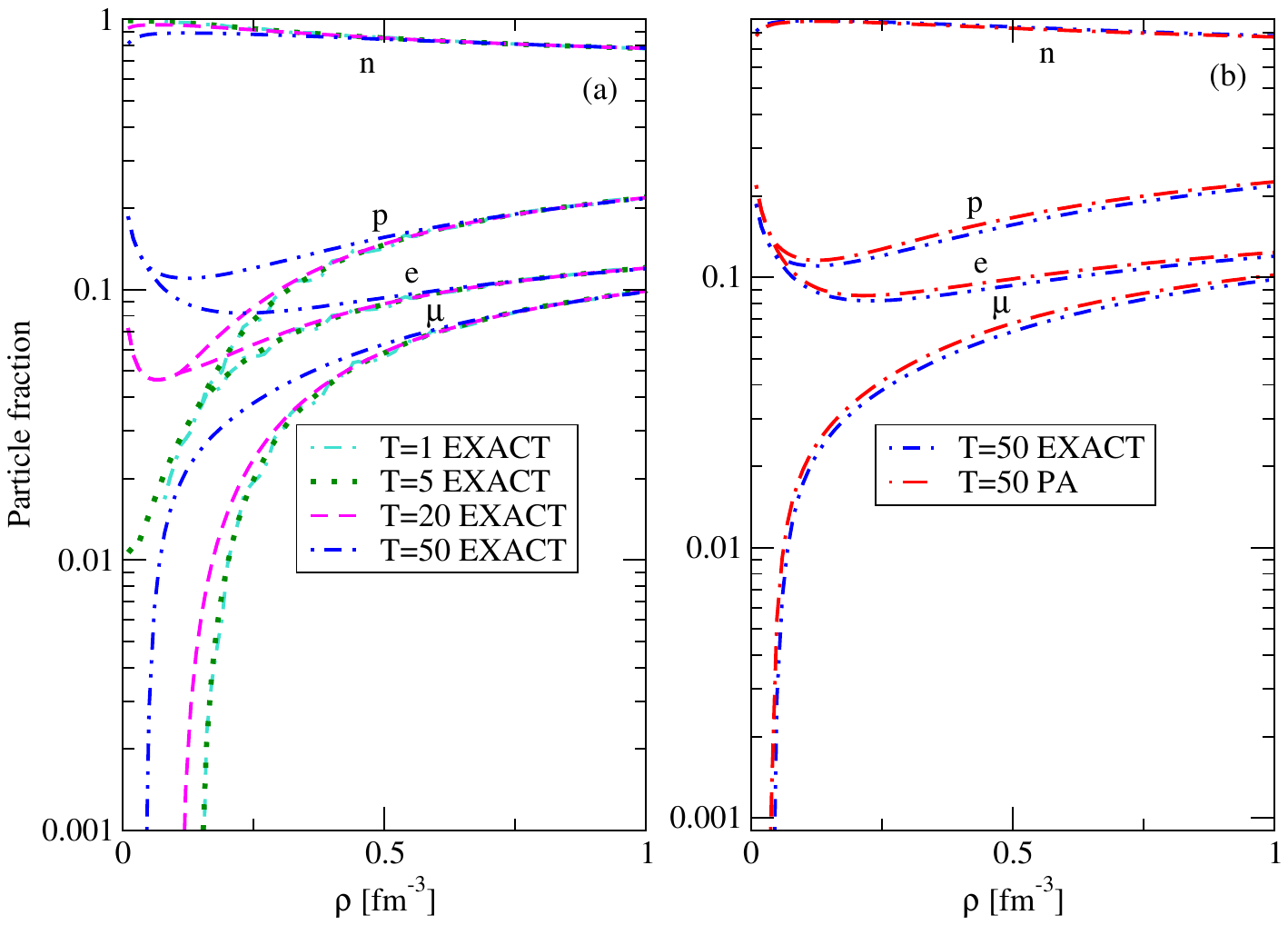}

		\caption{Panel (a): Particle fractions, $Y_n$, $Y_p$, $Y_e$, and $Y_\mu$, as a function of baryon density $\rho$ (fm$^{-3}$) for $T$=1,5,20,50 respectively; Panel(b): Comparison of particle fractions obtained using PA with the exact calculation results at $T$=50 MeV.}
		\label{fig2}
	\end{center}
\end{figure}

At a given $\rho$, the n,p FD distribution functions in NSM are evaluated for a given $T$ upon varying the isospin asymmetry $\delta=1-2Y_{p}$ subject to the simultaneous fulfillment of the $\beta$-equilibrium and charge neutrality conditions of Eq.(\ref{eq91}). The
$n,p,e$, and $\mu$ particle fractions in the hot NSM thus obtained are
shown in panel (a) of Figure \ref{fig2} as a function of density $\rho$ for $T$=1,5,20 and 50 MeV.
Upto $T$=5 MeV, the particle fractions remain almost the same as that of the zero-temperature results. But as $T$ increases the influence of temperature becomes evident, more prominently in the lower density range, where the neutron fractions decreases and simultaneously the proton, electron and muon fractions increases subject to the charge neutrality condition. The muon production has a threshold corresponding to its rest mass energy, $m_{\mu}c^{2}$ = 105.658 MeV, and density corresponding to this
threshold condition shifts to the lower density value as $T$ increases that can be seen from panel (a) of Fig.\ref{fig2}.
 As we go to higher densities, the influence of $T$ goes on gradually moderating ultimately merging all the curves to one,
for densities above $\approx$ 0.6 fm$^{-3}$.
Similar conclusion is also found in the work in Ref.\cite{Moustakidis2021} where MDI interaction is used. By comparing the SEI proton fractions data with the corresponding MDI predictions, it is found that both results are qualitatively similar, but with larger values in the MDI case.
 This is because MDI has a stiffer density dependence of the symmetry energy whose $L$-value is larger than the SEI one, while the $K_{sym}$ is positive for MDI (0.016 MeV) and negative in the SEI case (-37.2 MeV)
We have also compared the particle fractions at $T$=0 with the results of the Brueckner-Bethe-Goldstone (BBG) formulation of Burgio and Schulze given in the first panel of Fig. 4 in Ref.\cite{Burgio2010} where the comparison is quantitatively good.
 In panel (b) of Fig.\ref{fig2} we compare the $n$, $p$, $e$, and $\mu$ fractions under PA and exact calculations at $T$=50 MeV.
 The particle fractions of PA compares well with the exact predictions over the whole density range, overestimating slightly by the former to the latter. We have also checked that this trend is also true at other temperatures.

\begin{figure}[t]
	\begin{center}
		\includegraphics[height=11cm,width=18cm]{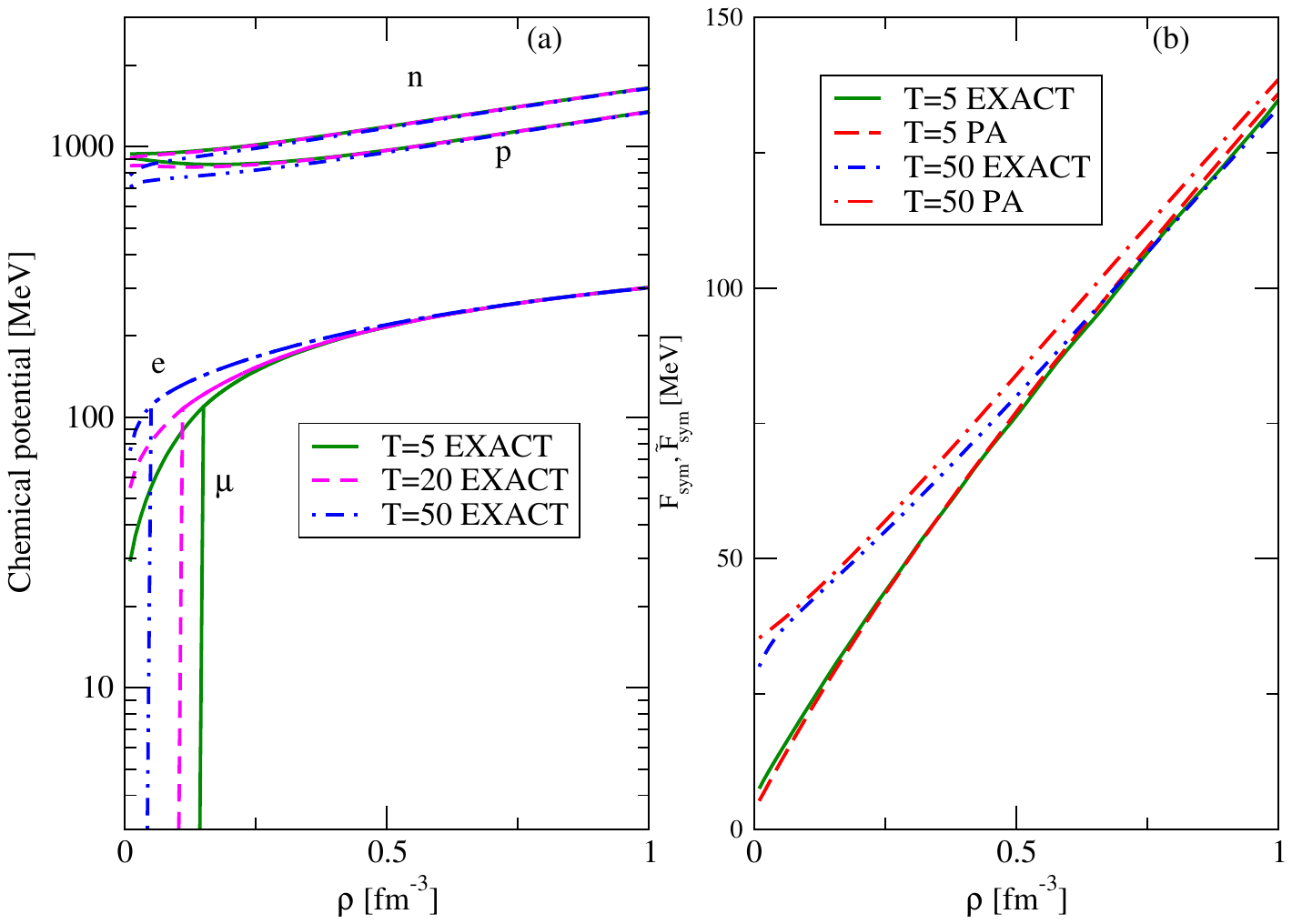}
		\caption{Panel (a): Chemical potentials of n and p obtained from the self-consistent exact calculation of NSM as a function of baryon density $(fm^{-3})$ at $T=5,20,50$. Panel (b): Comparison of free symmetry energy $F_{sym}(\rho,T)$ of PA with its exact calculation counterpart $\tilde{F}_{sym}(\rho,T)$ as a function of density at $T$=5 and 50 MeV.}
		\label{fig3}
	\end{center}
\end{figure}

 In panel (a) of Figure \ref{fig3} we display the exact $n$, $p$, $e$ and $\mu$ chemical potentials of the $\beta$-equilibrated hot NSM
as a function of density $\rho$ at $T$=5,20 and 50 MeV.
We can see that for the lowest considered temperature the difference ($\mu_n$-$\mu_p$) increases upto density
$\rho\approx$ 0.75 fm$^{-3}$ and thereafter maintains almost a constant value, while for the highest considered temperature this difference remains
practically constant for the whole range of considered densities.
This explains the merger of the
particle fractions curves for different $T$ in panel (a) of Fig.\ref{fig2} as $\rho$ increases.

 It has been verified in earlier works \cite{Bombaci1991,Prakash1997,Li2009} that at zero-temperature the PA given
by Eq.(\ref{eq9})
is a good approximation able to reproduce the exact $\beta$-equilibrium condition results of Eq.(\ref{eq8}) in NSM.
 In order to check the validity of the PA in the finite temperature domain, we have compared in panel (b) of Fig.\ref{fig3} the $F_{sym}(\rho,T)$
obtained under the PA from the calculation in PNM and SNM given in Eq.(\ref{eq10}) with the quantity $\tilde{F}_{sym}(\rho,T)$ = $(\mu_{n}-\mu_{p})/4(1-2Y_{p})$
, where $\mu_n$ and $\mu_p$ are obtained from the exact calculation in NSM, at temperatures $T$=5 and 50 MeV as a function of density ${\rho}$. This Figure shows that the PA given by Eqs.(\ref{eq9}) and (\ref{eq10}) is also valid
in the finite temperature domain. It implies that the higher order terms (4$^{th}$-order onwards) in the Taylor series expanded energy for ANM at finite $T$ have negligible contribution. The small higher values of $F_{sym}$ of PA over the exact data of $\tilde{F}_{sym}(\rho,T)$, explains the relatively larger particle fraction values of the PA, shown in panel (b) of Fig.\ref{fig2}, which can be understood from the relation in Eq.(\ref{eq9}).

\begin{figure}[t]
	\begin{center}
		\includegraphics[height=11cm,width=18cm]{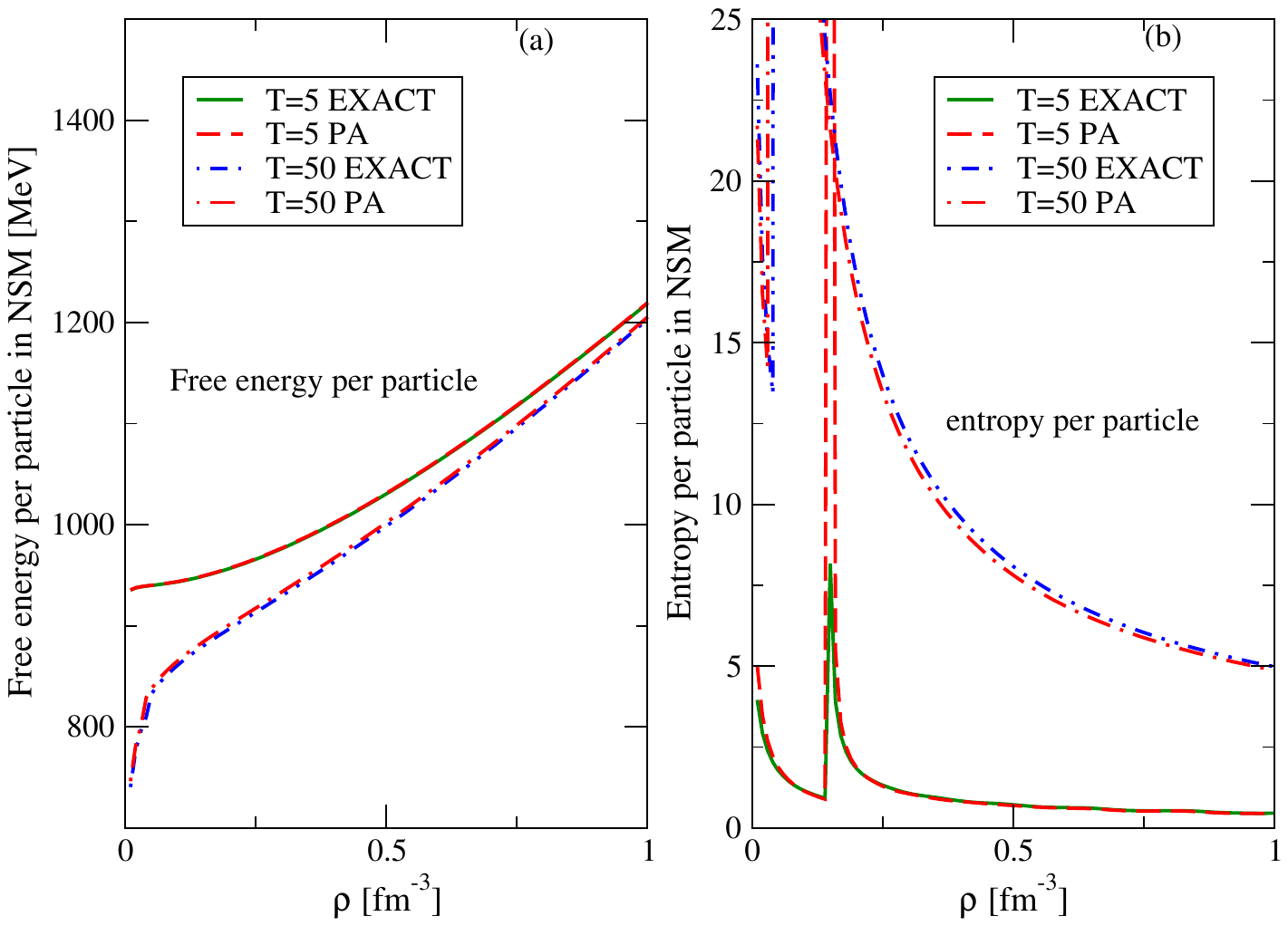}
		\caption{Panel (a): Free energy per particle $\bar{F}_{NSM}$ in (MeV), obtained under the exact and PA, is shown as a function of baryon density (fm$^{-3}$) at $T$=5 and 50 MeV in NSM. Panel (b): Entropy per particle $\bar{S}_{NSM}$ (in the unit of Boltzmann constant $k_B$), obtained under the exact and PA, in NSM as a function of baryon density (fm$^{-3}$) for $T$=5 and 50 MeV.}
		\label{fig4}
	\end{center}
\end{figure}

\begin{figure}[t]
	\begin{center}
		\includegraphics[height=11cm,width=18cm]{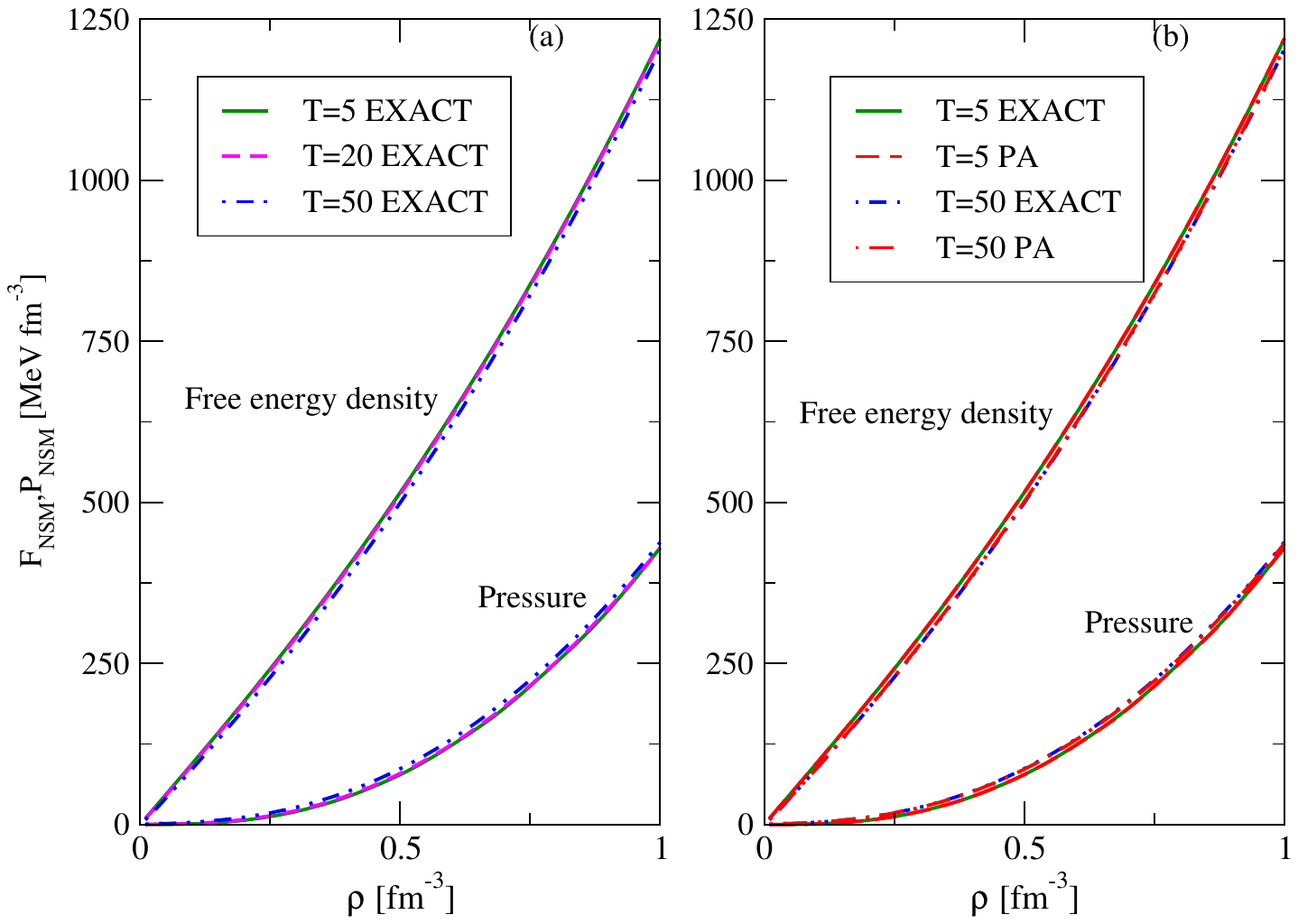}
		\caption{Panel (a): Total free energy density and pressure, under exact calculation, as a function of baryon density at $T$=5,20,50 MeV in the neutrino-free NSM. Panel (b): The PA results for free energy density and pressure in NSM compared with the exact results at $T$=5 and 50 MeV. }
		\label{fig5}
	\end{center}
\end{figure}

 The free energy per nucleon in NSM is given by $\bar{F}_{NSM}$=$\frac{F_{NSM}}{\rho}$, where $F_{NSM}$ is the
free energy density given by the first equation (\ref{eq11}), which comprises of the nucleonic and leptonic parts. The nucleonic
free energy density is computed as
$F_N$=$H_N-TS_N$, where $H_N$ and $S_N=S_n+S_p$ are the nucleonic energy and entropy densities, respectively.
The results for $\bar{F}_{NSM}$ obtained under the exact evaluation
are shown as a function of $\rho$ for $T$=5 and 50 MeV in panel (a) of Figure \ref{fig4}. The influence of $T$ on $\bar{F}_{NSM}$ is prominent in the low density region, where the decrease in $\bar{F}_{NSM}$ is larger for higher temperature $T$ owing
to the higher value of entropy. As the density increases this decreasing trends gets moderated, and the curves for all the $T$ gradually approach the asymptotic value of the zero-temperature limit. This is also the finding in Ref.\cite{Moustakidis2021} for the MDI interaction. The larger decrease in $\bar{F}_{NSM}$ at higher $T$ is due to the higher value of entropy. The entropy per particle in NSM, $\bar{S}_{NSM}$=$\frac{S_{NSM}}{\rho}$, is computed from the expression,
\begin{eqnarray}
\bar{S}_{NSM} =\bar{S}_{N} + \bar{S}_{e}+ \bar{S}_{\mu},
	\label{eq18}
\end{eqnarray}
where the nucleonic part of the entropy density $S_{N} =(S_{n}+S_{p})$, $S_{n(p)}$ is calculated using Eq.(\ref{eq13})
with the FD functions obtained from the exact evaluation. The leptonic entropy densities, $S_{e}$ and $S_{\mu}$, are also calculated
with the same expression (\ref{eq13}) but with the nucleonic FD functions replaced by the leptonic ones. The total entropy per particles
at temperatures of 5 and 50 MeV are displayed in panel (b) of Fig.\ref{fig4}.
 In general, $\bar{S}_{NSM}$ is large at lower value of density, having higher magnitude for larger values of temperature and decreasing
with increasing density. The muon production is marked by a resonance peak of Breit-Wigner type at each temperature $T$ 
in the corresponding curve at the threshold density, which shifts to lower density values when temperature increases. Entropy being
the measure of disorderness of the system, it increases sharply on the appearance of a new particle in the system. This explains the decreasing
trend of the free energy per particle in NSM
 with increasing temperature in panel (a) of Fig.\ref{fig4}. The pressure in NSM, $P_{NSM}$, is given by the second equation of
(\ref{eq11}), which is evaluated under exact calculation using the Eqs.(\ref{eq14}) and (\ref{eq15}) for nucleonic and leptonic parts, respectively.
The results of free energy density, $F_{NSM}$ and pressure, $P_{NSM}$ in NSM, under the exact calculation, are shown as a function of density at
$T$=5, 20, 50 MeV in panel (a) of Figure \ref{fig5}. It is evident from the results of $F_{NSM}$ and $P_{NSM}$ at different $T$ in this
Fig.\ref{fig5} that effect of temperature on the EoS of NSM is minimal. In Ref.\cite{Moustakidis2021}, similar results are found where MDI is used.

 We check now the validity of the PA so far as the thermodynamical quantities in hot NSM are concerned.
Under the PA, $H_N=H_{SNM}+(1-2Y_p)^2(H_{PNM}-H_{SNM})$, $P_N = P_{SNM}+(1-2Y_p)^2(P_{PNM}-P_{SNM})$ and $S_N=S_{SNM}+(1-2Y_p)^2(S_{PNM}-S_{SNM})$, where $H_{SNM}$, $H_{PNM}$; $P_{PNM}$, $P_{SNM}$
and $S_{SNM}$, $S_{PNM}$ are the energy, pressure and entropy densities in SNM and PNM, respectively.
From panel (a) Fig.\ref{fig4} we can see that the exact and PA results for free energy per particle are practically the same at all temperatures.
The comparison between the exact and PA entropy per particle as a function of the density, displayed in panel (b) of Fig.\ref{fig4}, show that
the exact and PA results are in good agreement over the whole range of density at all temperatures considered.
In panel (b) of Fig.\ref{fig5}, we have also compared the exact and PA results for the total free energy and pressure at $T$=5 and 50 MeV. Both,
the exact and PA, results are quite similar. Thus, in the case of our SEI interaction, the PA is a powerful tool in hot NSM calculations,
 able to reproduce the exact particle fractions and thermodynamical quantities with very high precision. However, the advantage of the exact
evaluation over the PA is that one gets directly the n,p chemical potentials and the respective FD distribution functions of the nucleons and
leptons in NSM, which is not the case in PA. The studies in the followings shall be made under the exact calculation.


\begin{figure}[t]
	\begin{center}
		\includegraphics[height=11cm,width=18cm]{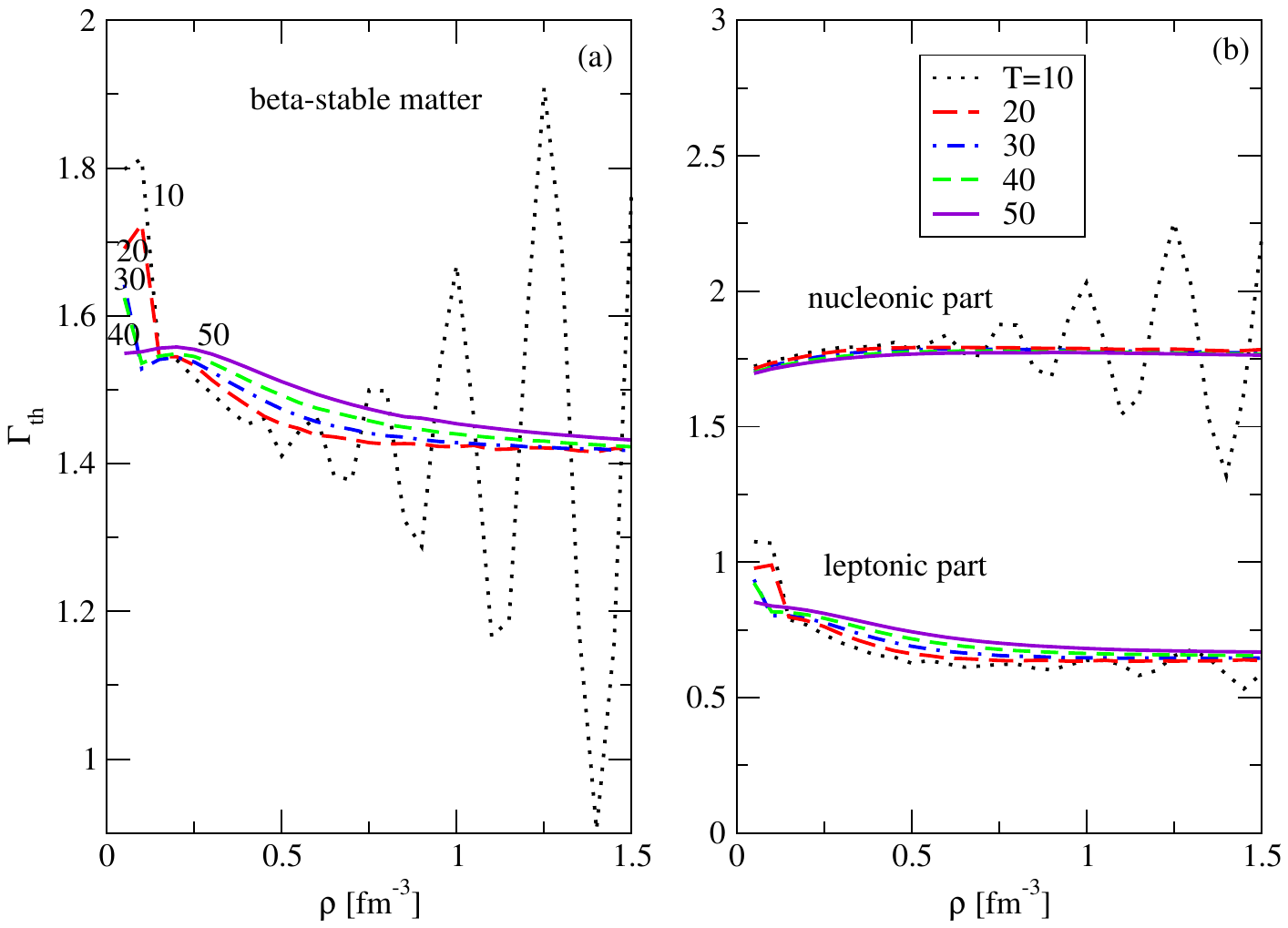}

		\caption{(Panel (a)): Thermal index $\Gamma_{th}$ as a function of baryon density $(fm^{-3})$ for $T$=1,5,20,50 MeV in NSM. (Panel (b)): Thermal index for nucleonic and leptonic parts of NSM as a function of baryon density at $T$=10,20,30,40 and 50 MeV. The legends for both panels are the same.}
		\label{fig6}
	\end{center}
\end{figure}

  The crucial role of the thermal part of the pressure has been ascertained from the NSs merger simulation studies where its influences on the merger remnant and frequency of the emitted GWs have been found \cite{Bauswein2010,Sekiguchi2011,Paschalidis2012}. The thermal contribution in the simulation studies is often taken into account in term of the $\Gamma$-law prescription \cite{Bauswein2010,Kiuchi2014} where a constant value for the thermal index parameter, $\Gamma_{th}$, is used. The $\Gamma_{th}$ is defined as,
	\begin{eqnarray}
\Gamma_{th} = 1+ \frac{P^{th}_{NSM}}{H^{th}_{NSM}} = 1+ \frac {P^{N}_{NSM}(T,\rho)-P^{N}_{NSM}(T=0,\rho)} { H^{N}_{NSM}(T,\rho)-H^{N}_{NSM}(T=0,\rho)}\nonumber\\
               + \frac {P^{l}_{NSM}(T,\rho)-P^{l}_{NSM}(T=0,\rho)} { H^{l}_{NSM}(T,\rho)-H^{l}_{NSM}(T=0,\rho)}
          \label{eq19}
\end{eqnarray}
where, $P^{i}_{NSM}$ and $H^{i}_{NSM}$, $i$=$N$ and $l$, are the pressure and energy density of the nucleonic and leptonic components, respectively, in NSM. The thermal evolutions in these two parts will be different as they are governed by different force laws and need to be considered separately.
 A constant value for the thermal index in the range 1.5--2 is commonly used, ignoring its density dependence. For an ideal fluid EoS, the thermal index is constant and takes the value 5/3 for nonrelativistic fermions and 4/3 for ultra-relativistic fermions.
The results of $\Gamma_{th}$, in Eq.(\ref{eq19}), computed for the SEI EoS at $T$=10,20,30,40 and 50 MeV is shown as a function of density $\rho$ in panel (a) of Figure \ref{fig6}.
At low $T$ we found a highly oscillating behaviour for $\Gamma_{th}$, which shifts to higher density with much reduced amplitude as $T$ increases. This may be assigned to the fact that at low value of $T$, where the Fermi kinetic energies of zero-temperature n,p distributions are larger than the energy $k_{B}T$, the distribution functions remains practically the same as that for $T$=0. This can be seen from Fig.\ref{fig1} where, for a given temperature, the depletion of the nucleons from Fermi level and the occupancy of the higher excited states decreases when the density increases.
The curves of $\Gamma_{th}$ of NSM for the different $T$ shown in panel (a) of Fig.\ref{fig6} starts from different values at low density below $0.05 fm^{-3}$, but approaching to a constant value $\Gamma_{th}$$\sim$1.5 as density increases showing a small decreasing trend. In panel (b) of Fig.\ref{fig6} we have shown the thermal index for the nucleonic part and the leptonic part at the same temperatures, $T$=10,20,30,40 and 50 MeV. The thermal index for the nucleonic part shows almost a perfect constant behaviour about the value $\Gamma_{th}$$\sim$1.78 for all temperatures. The thermal index for leptonic part also shows nearly a constant behaviour with a relatively wider spread for different $T$ and having a small decreasing trend which has been reflected in the NSM results of panel (a). The wider differences amongst the curves of different $T$ in panel (a) at the low density below 0.2 fm$^{-3}$ is due to the resonances in the leptonic curves at the threshold density of muon production. So, the constant value $\Gamma_{th}$ used in the core collapsing PNS and BNSM simulation studies strictly holds for the nucleonic component, because the $\Gamma_{th}$ for NSM as a whole shows a weak $T$ and $\rho$ dependence.

\subsection{Neutrino-trapped hot NSM }

  In the case where neutrinos are trapped and therefore are a constituent of the system, and therefore will contribute
to the thermodynamics and EoS of the matter.
  Now the $\beta$-equilibrium condition will be achieved according to the Eq.(\ref{eq8}) (which is rewritten in Eq.(\ref{eq16})). The neutrino-trapped $\beta$-equilibrated matter is encountered in the core collapse supernovae matter forming PNS as well as in the highly oscillating metastable remnant in the BNSM in delayed collapse, where the matter can be considered as in isoentropic condition.
In both scenarios, the large temperatures resulting from the collapse or merger dynamics produce a copious amount of neutrinos, which initially form a trapped neutrino gas and diffuse out over the diffusion timescales ($\sim$ seconds).
Numerical simulations show that in both scenarios the bulk of matter is characterized by a quasi-uniform, low entropy-per-baryon
profile $\bar{S}$ = 1 - 3 (in units of $k_B$), which decreases only over the cooling timescale because of
neutrino emission \cite{Lattimer1986,Prakash1997,Pons1999,Fischer2010,Hudepohl2010,Roberts2012,Kastaun2015,Perego2019}.
 In the neutrino-trapped scenario the lepton fraction for each kind of flavour is defined as $Y_{L_{l}}=Y_{l}+Y_{\nu_l}$ ($l=e, \mu$) where
$Y_l=\rho_l/\rho$ and  $Y_{\nu_l}=\rho_{\nu_l}/\rho$. Due to the lepton number conservation, $Y_{L_{l}}$ has to be fixed separately for
each type of lepton. In the merged matter formed in the BNSM, the isospin asymmetry is high with lepton fractions $Y_{L_{l}}\sim 0.1$, for both electrons
and muons, whereas, the supernovae matter is more isospin symmetric, $Y_{L_{l}}\sim 0.3-0.35$ and the leptonic population is
dominantly electronic \cite{Prakash1997,Bombaci2021,Malfatti2019,Alford2019}.
\begin{figure}[t]
	\begin{center}
		\includegraphics[height=11cm,width=18cm]{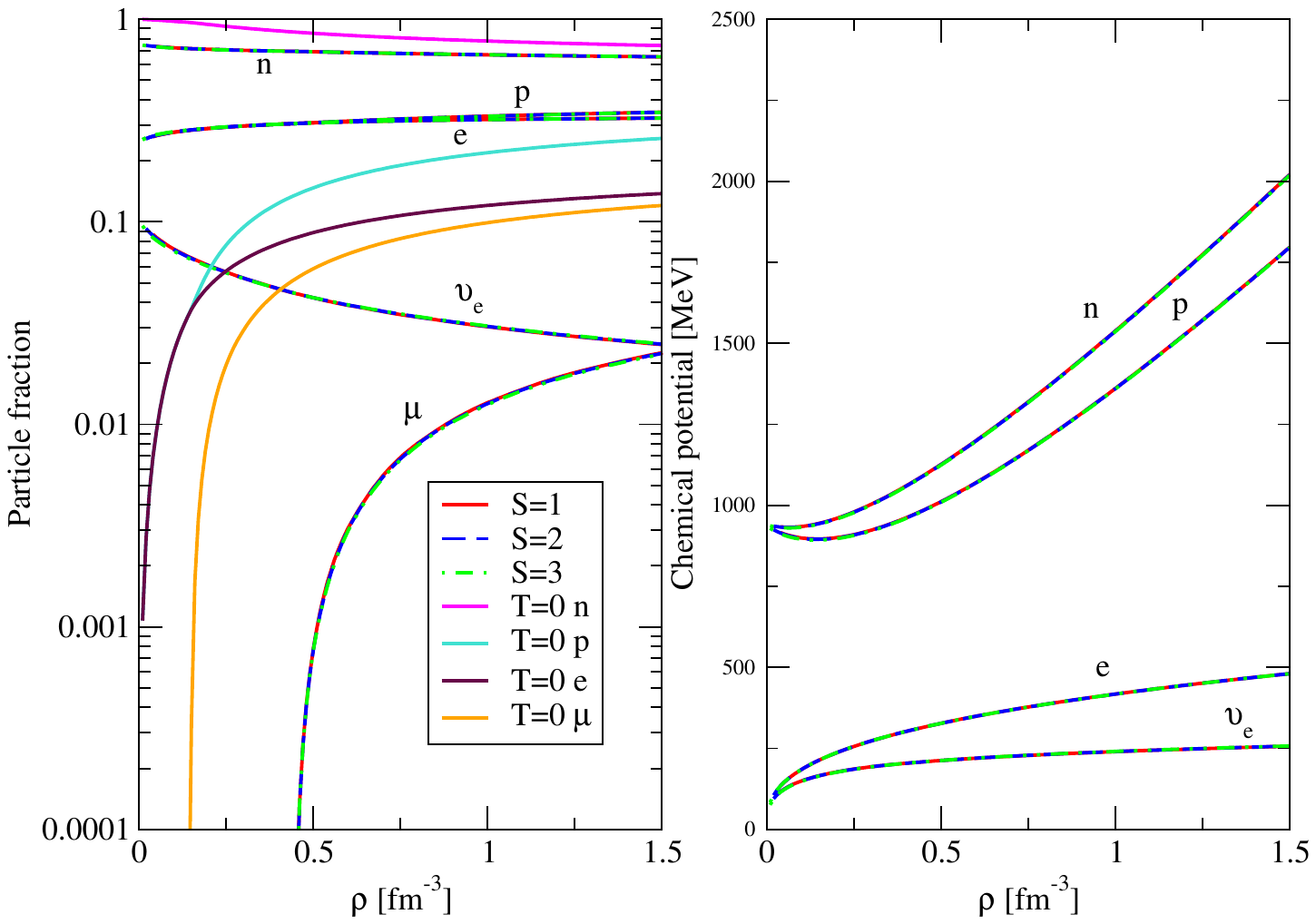}
		\caption{(Panel (a)): The particles fractions, $Y_i, i=n, p, e, \nu_e$, in the isoentropic matter for $\bar{S}$=1, 2 and 3 for the SEI EoS $\gamma$=2/3, $L$=75 MeV are shown as a function of density $\rho$. The data of $Y_i, i=n, p, e, \mu$ for $T$=0, neutrino free matter are also given for comparison.(Panel (b)): The chemical potentials, $\mu_i, i=n, p, e, \nu_e$, obtained in the three cases, $\bar{S}$=1, 2 and 3, are shown as a function of density $\rho$.}
		\label{fig8}
	\end{center}		
			 \end{figure}
\begin{figure}[t]
	\begin{center}
		\includegraphics[height=11cm,width=18cm]{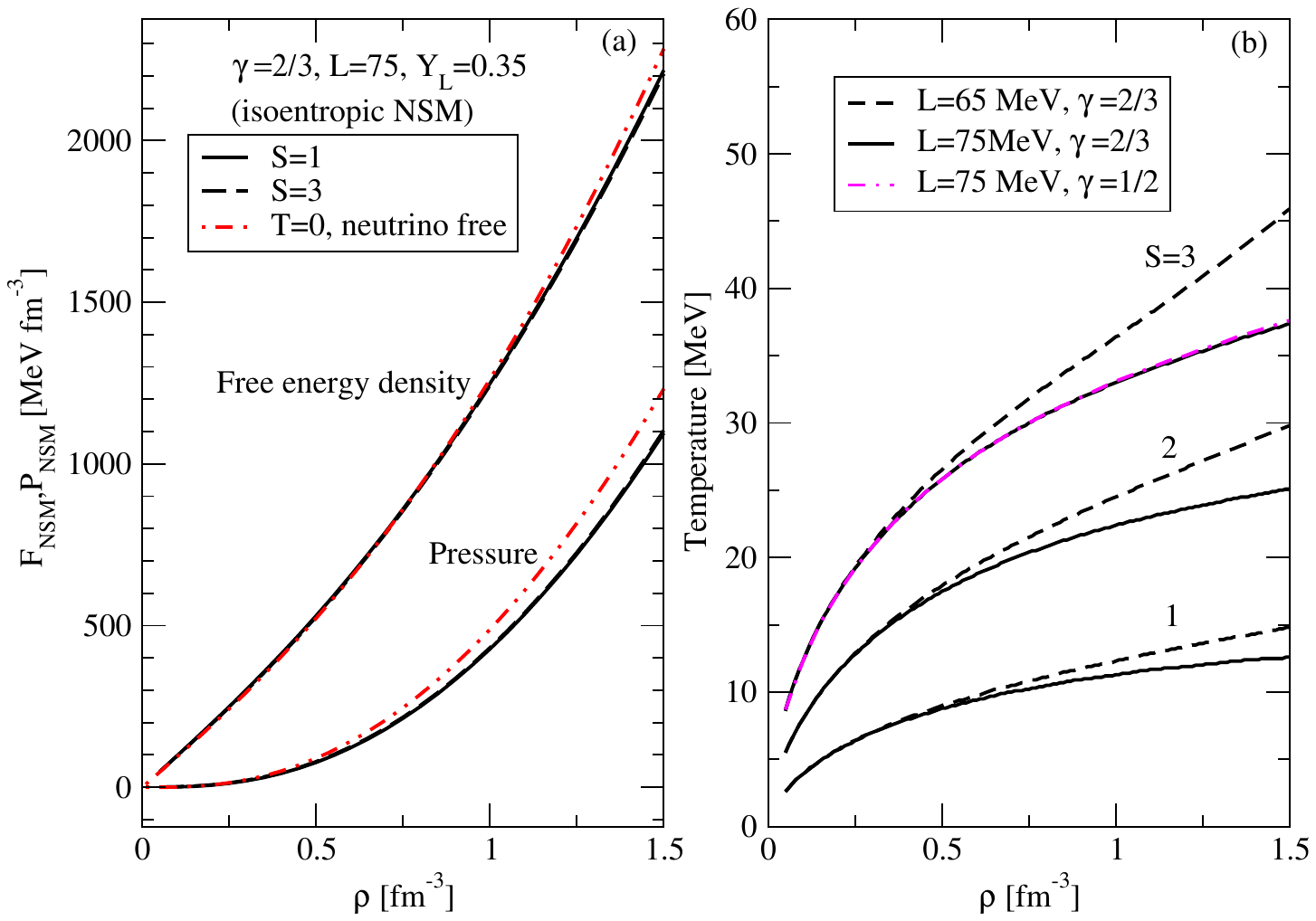}
		\caption{(Panel (a)): The EoSs of the three isoentropic cases of $\bar{S}$=1, 2 and 3 where the free energy density $F_{NSM}$ and pressure $P_{NSM}$ are shown as a function of density $\rho$. The neutrino free $T$=0 results are also given. (Panel (b)): Isentropes in the $\rho$-$T$ plane for three values of entropy per particles, $\bar{S}$=1, 2 and 3 for the SEI EoS $\gamma$=2/3 and two slope values L=65 and 75 MeV. The data for the isentrope $\bar{S}$=3 for the SEI EoS $\gamma$=1/2, $L$=75 MeV is also given.}
		\label{fig9}
	\end{center}		
			 \end{figure}
The equilibrium particle fractions are calculated for the PNS supernovae neutrino-trapped matter taking into account only electrons and electronic neutrinos and considering three constant values of entropy per particle, $\bar{S}$=1, 2 and 3 (in the units of $k_B$) in isoentropic NSM by solving the $\beta$-equilibrium condition (\ref{eq16}) using for the SEI-Y EoS ($\gamma$=2/3, $L$=75 MeV).
 The particle fractions, $Y_i$, and chemical potentials,
$\mu_i$, $i=n, p, e, \nu_e$, in the PNS scenario are shown in panels (a) and (b) of Figure \ref{fig8}, respectively, as functions
of density $\rho$ for three constant values of the entropy per particle, $\bar{S}$=1, 2 and 3 in isoentropic NSM. The differences in the particle fractions for the three entropies per particle considered in the figure are small and indistinguishable at the scale of the figure. A similar situation is found for the chemical potentials.
Similar results are found for the particle fractions in the three lower panels of Fig. 4 of Ref.\cite{Burgio2010}, which were obtained in the BBG formulation of isoentropic matter for $\bar{S}$=0, 1, and 2, as we have verified. This shows that the composition of the matter under isoentropic condition is insensitive to the entropy of the system. However, comparing the particle fractions in the trapped matter with those of the neutrino-free matter at zero temperature, which are also given in panel (a) of Fig.\ref{fig8} for comparison, it can be seen that the electron and proton fractions in the trapped matter are higher than those at $T$=0 MeV. This is because the electron chemical potential $\mu_e$ in the trapped matter increases by an amount of $\mu_{\nu_{e}}$ resulting in higher electron fraction, and thereby the proton fraction is increased as a consequence of charge neutrality. The muon fraction in the neutrino-trapped matter also decreases and the muon production threshold density shifts to high density.

 The EoSs of the isoentropic matter for $\bar{S}$=1 and 3 are shown in panel (a) of Figure \ref{fig9}, where the free energy density and pressure are shown as a function of $\rho$. The differences between the results for $\bar{S}$=1 and 3 in both the quantities are small and indistinguishable at the scale of the figure. The zero temperature, $T$=0, neutrino free EoS is also shown in the same panel (a) of Fig.\ref{fig9}. The EoS of neutrino-trapped matter is relatively softer than the neutrino-free matter, as found in other model calculations \cite{Prakash1997,Bombaci2021}.

In panel (b) of Fig.\ref{fig9}, we display the isentropes in the $\rho$-$T$ plane for the entropy values of $\bar{S}$=1, 2 and 3, which show a remarkable entropy dependence. The temperature values of an isentrope for higher $\bar{S}$ is larger. This is understood as, entropy being the measure of disorderness, matter at a given density $\rho$ requires larger temperature $T$ to remain at higher $\bar{S}$. The temperature profile of an isentrope of given $\bar{S}$ has an increasing trend with rise in density $\rho$ which is stiffer in the lower density region.
 The density dependence of symmetry energy mostly decides the particle fractions in NSM. In order to investigate the influence of the stiffness of the symmetry energy on the isentropes, we have computed the isoentropes at $\bar{S}$=1, 2 and 3 with the SEI EoS where the slope parameter $L$=65 MeV, instead of the value 75 MeV (see Table {\ref{tab1}}). Within the SEI model a change of the $L$-value leaves the saturation properties of SNM and the $n,p$-effective mass
splitting in ANM invariant, but predicts a relatively softer density dependence of the symmetry energy. For the SEI EoS with
$L$=65 MeV we have verified that the changes in the particle fractions, chemical potentials and the EoSs are rather small compared to those
predicted by the SEI model with $L$=75 MeV, which are
shown in the Figs.\ref{fig8} and \ref{fig9}. However, there is a  noticeable difference between the isentropes of $L$= 75 and 65 MeV in the three cases of $\bar{S}$=1, 2, 3, shown in panel (b) of Fig.\ref{fig9}. The isentrope for lower $L$ value of 65 MeV results into stiffer temperature profile in all the three cases of $\bar{S}$=1, 2, 3. This is because a softer EoS predicts relatively smaller particle fractions in a given volume
and, therefore, a higher $T$ is necessary to maintain the system at given $\bar{S}$. We shall discuss further on this aspect in the context of NS in the following.
The influence of incompressibility $K$, which is the stiffness of the symmetric matter, on the isentrope has been examined by computing the isentrope for $\bar{S}$=3 using the SEI EoS having $\gamma$=1/2 ($K$=237 MeV) but the same slope of the symmetry energy $L$=75 MeV. The result of the calculation is shown in panel (b) of Fig.\ref{fig9}. We see that the isentropes in this case are very similar to those predicted by SEI EoS with $\gamma$=2/3 and $L$=75 MeV EoS. From these analyses we conclude that the temperature profile in isoentropic matter is sensitive to asymmetric stiffness, i.e., $L$-value, whereas, its dependence on the symmetric stiffness, i.e., on $K$-value, is not significant.

\begin{figure}[t]
	\begin{center}
		\includegraphics[height=11cm,width=18cm]{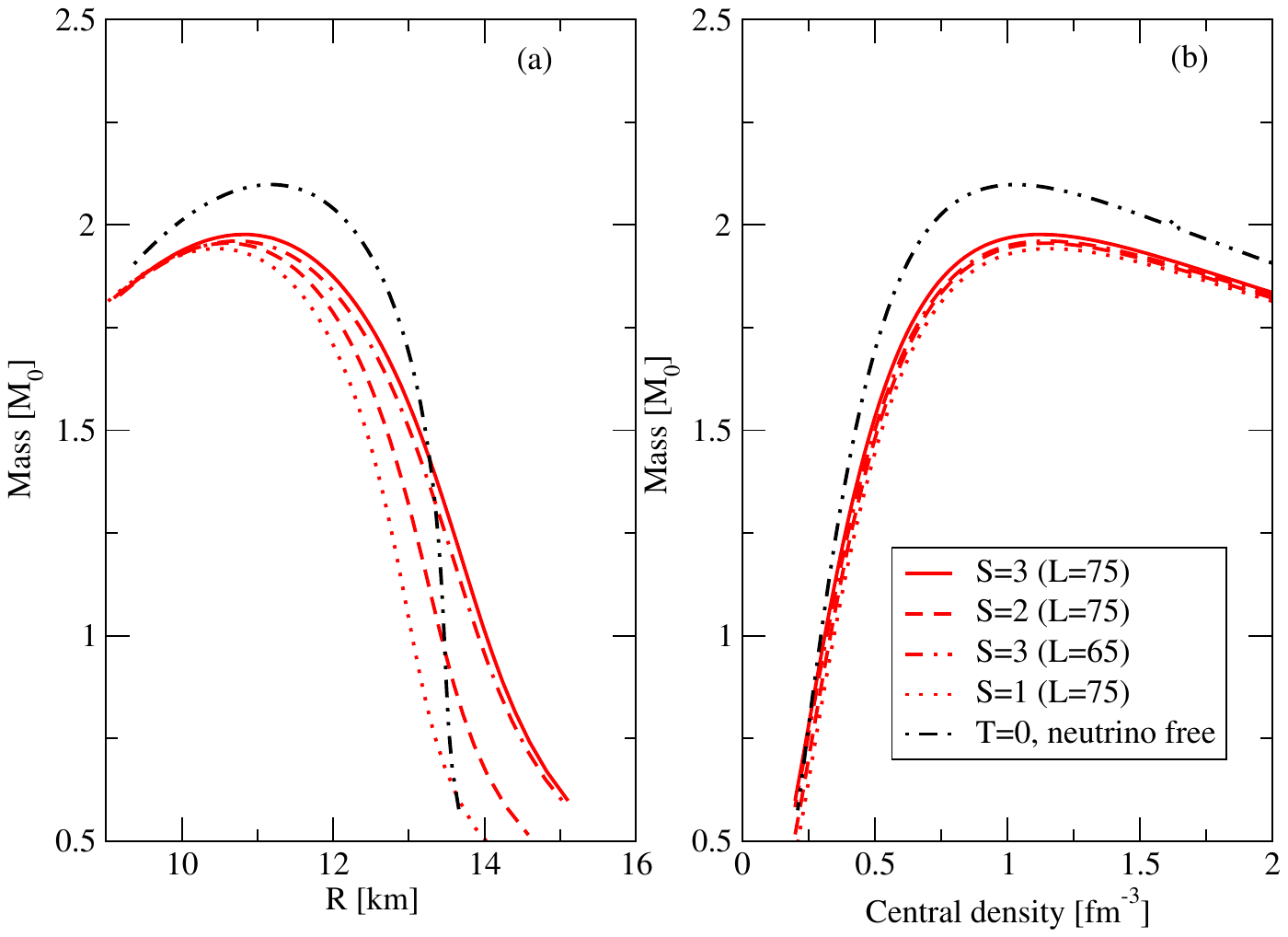}
		\caption{(Panel (a)): Neutron star mass-radius relation for the isoentropic EoSs of $\bar{S}$=1, 2, 3 for the SEI interaction $\gamma$=2/3, $L$=75 MeV. The results for $L$=65 MeV of $\bar{S}$=3 together with the $T$=0 neutrino-free SEI EoSs are also shown. (Panel (b)): The NS mass-central density relation shown for the same EoSs of panel (a). The legends are same for both the panels. }
		\label{fig10}
	\end{center}		
			 \end{figure}
The NS mass-radius relation calculated for the three neutrino-trapped isoentropic EoSs corresponding to $\bar{S}$=1, 2, 3 is done by solving the Tolman-Oppenheimer-Volkoff (TOV) equation using our SEI-Y EoS having $L$=75 MeV, where we have used for the crust region the BPS-BBP EoS \cite{Baym1971a,Baym1971b} upto density $\rho$=0.058 fm$^{-3}$ and our EoSs results thereafter. The gravitational mass (GM)
$M_{G}$ (in solar mass units), computed with the three isentropic EoS analyzed in this work, is displayed as a function of the radius
of the star $R$ and its central density $\rho_c$ in panels (a) and (b) of Figure \ref{fig10}, respectively. The results corresponding to the $\bar{S}$=3 isentrope computed with the SEI EoS $L$=65 MeV together with the zero-temperature
results in the neutrino-free case are also shown in panels (a) and (b) of Fig.\ref{fig10}.
The maximum mass $M_{G}$ predicted by the $T$=0 neutrino-free EoS of SEI is 2.098 $M_{\odot}$, which satisfies the maximum mass constraint
of 2.08$^{+07}_{-07}$ $M_{\odot}$ measured in the  PSR J0740+6620 \cite{Fonseca2021}. The $T$=0 neutrino-free EoS of SEI also conforms
the 1.4 $M_{\odot}$ NS radius constraint, $R_{1.4}$=12.45$\pm$0.65 km, extracted from the analysis of PSR J0740+6620 data \cite{Miller2021} and
the limit R$_{1.4}$=11.9$\pm$1.4 km, ascertained by LIGO-Virgo collaboration \cite{Abbott2018}. The maximum mass predicted by the three isoentropic EoSs of $\bar{S}$=1, 2, 3 are 1.942, 1.956 and 1.977 $M_{\odot}$, respectively. These masses show an increasing
trend with increasing entropy, as it can be seen from panel (a) of Fig.\ref{fig10}. The NS maximum mass predicted by the neutrino-trapped
isoentropic EoSs are smaller than the maximum mass of $T$=0 neutrino-free EoS case. This is in agreement with the trend found in the study using BL
microscopic EoS in Ref.\cite{Bombaci2021}. The softening of the EoS of neutrino-trapped matter due to the increase in electron and proton fractions
owing to the additional electron neutrino chemical potential is discussed in the foregoing discussion.

The NS baryonic mass, $M_B$, predicted by the different EoSs considered in this work is computed by integrating the baryon number density in the TOV solution, which gives the total baryon number $N_{ns}$ in the NS, and the corresponding baryonic mass, defined as $M_B$=$N_{ns}m_u$, where $m_u$=1.6605 10$^{-24}$g is the atomic mass unit.
 The baryonic masses $M_B$ in solar mass units corresponding
to the maximum gravitational mass $M_{G}$ of the NS predicted by each EoS used in this work is reported in Table (\ref{tab2}).
From this Table we can also see that
the baryonic mass in the maximum gravitational mass NS computed with the $L$=75 MeV EoS is higher than that of the $L$=65 MeV EoS. The baryonic masses of the 1.4 $M_{\odot}$ mass NSs predicted by the EoSs of $L$=75 and 65 MeV are 1.538 and 1.530 $M_{\odot}$, respectively. Thus a lower $T$ will be required in the former where $L$ is higher to maintain the system at the same entropic condition than the later of lower $L$-value EoS which has less number of nucleons.
The compactness, $C=GM_{G}/{Rc^2}$, where $G$ is the gravitational constant and $c$ is the velocity of light, of the maximum mass NSs of
the neutrino-trapped and $T$=0, neutrino-free EoSs are given in Table \ref{tab2}. The compactness of the maximum mass NSs with a
neutrino-trapped scenario remain almost constant when the entropy per baryon varies from $\bar{S}$=1 to $\bar{S}$=3, as can be seen in the same Table.
 However, the compactness of the maximum mass NS decreases if the $L$-value of the EoS decreases as it can also be seen in the same Table \ref{tab2}. This EoS dependence might have observable effect in the simulation studies of core-collapse supernovae and protoneutron stars, as well as, in BNSM.

\begin{figure}[t]
	\begin{center}
		\includegraphics[height=11cm,width=18cm]{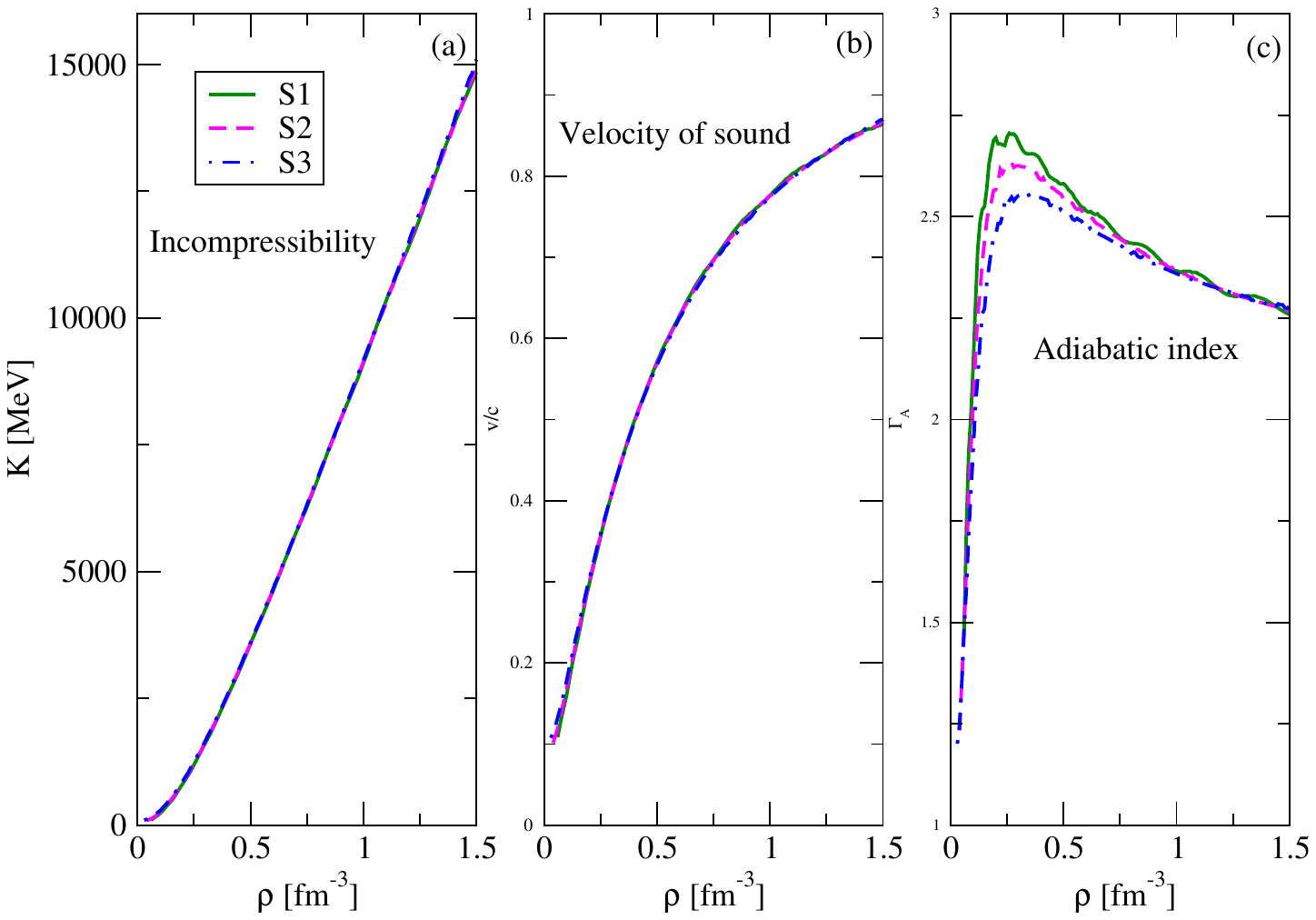}
		\caption{(Panel (a)): Incompressibility $K$ as a function of baryon density (fm$^{-3}$), (Panel (b)): Velocity of sound as a function of baryon density and (Panel(c)): Adiabatic index $\Gamma_{A}$ as a function of baryon density in neutrino-trapped isoentropic NSM for the three EoSs corresponding to $\bar{S}$=1, 2, 3.}
		\label{fig7}
	\end{center}		
			 \end{figure}
    \begin{table}[h]
    \caption{Neutron star central density $\rho_{c}$, radius R, maximum mass $M{_G}$, baryonic mass $M_{B}$, radius of 1.4 M$_\odot$ mass NS R$_{1.4}$, compactness $C^{max}$ of maximum mass NS for both isoentropic $\bar{S}$=1, 2, 3, and $T$=0 neutrino-free EoSs of SEI ($\gamma$=2/3, $L$=75, 65 MeV). }
    	\centering
    	   \begin{scriptsize}
    		\begin{tabular}{ccccccc}
    			\hline
    			EoS & $\rho_c (fm^{-3})$ &$R (km)$ & $M{_G} (M_{\odot})$ & $M_B (M_{\odot})$ & $R_{1.4}(km)$ & $C^{max}$\\

    			\hline
    			SEI-Y($\gamma$=2/3,$L$=75 MeV,$S$=3)&1.13&10.536&1.977&2.274&12.467&0.2768\\
    			SEI-Y($\gamma=2/3$,$L$=75 MeV,$S$=2)&1.15&10.60&1.956&2.236&12.90&0.2722\\
    			SEI-Y($\gamma=2/3$,$L$=75 MeV,$S$=1)&1.16&10.472&1.942&2.211&12.590&0.2736\\
					SEI($\gamma=2/3$,$L$=65 MeV,$S$=3)&1.15&10.742&1.961&2.257&13.194&0.2695\\
					SEI($\gamma=2/3$,$L$=75 MeV,$T$=0,$\nu_e$=0)&1.03&11.185&2.098&2.470&13.29&0.2767\\
    			\hline
    			\label{tab2}
    		\end{tabular}
   	\end{scriptsize}
    \end{table}
	%
	%
The incompressibility of the hot isoentropic NSM is calculated from the expression given by,
 \begin{eqnarray}
 	K(\rho,Y_{p},T)= 9\left({\frac{\partial P_{NSM}(\rho,Y_{p},T)}{\partial\rho}}\right)_{\bar{S}}=\nonumber \\
9\left({\rho}^{2} {\frac{\partial^{2} e_{NSM}(\rho,Y_{p},T)}{\partial\rho^{2}}}
     + 18\rho \frac{\partial e_{NSM}(\rho,Y_{p},T)} {\partial\rho}\right)_{\bar{S}},
\label{eq20}
\end{eqnarray}																												
 where, $P_{NSM}(\rho,Y_{p},T)$ and $e_{NSM}(\rho,Y_{p},T)$ are the pressure and energy per nucleon in hot neutrino-trapped NSM at constant entropy
$\bar{S}$. The incompressibility value at the core density of the maximum mass configuration at finite $T$ is crucial in deciding whether the two NSs merger will undergo a prompt/delayed collapse \cite{Perego2022}.
 The velocity of sound $v$ in NSM in the unit of velocity of light $c$ is calculated from the relation,
\begin{eqnarray}
\frac{v}{c} = \left(\frac{\partial P_{NSM}}{\partial H_{NSM}}\right)_{\bar{S}}^{1/2},
\label{eq21}
\end{eqnarray}
where $P_{NSM}$ and $H_{NSM}$ are the total pressure and energy density of the hot isoentropic NSM. The results for the incompressibility $K$ and velocity of sound $\frac{v}{c}$ in hot isoentropic NSM are shown in panels (a) and (b) of Figure \ref{fig7}, respectively, as a function of density $\rho$ for the three EoSs corresponding to $\bar{S}$=1, 2, 3.
Both $K$ and $\frac{v}{c}$ are increasing functions of density where the influence of temperature profile inside NSM corresponding to the three $\bar{S}$ values is minimal. At $\rho=1.5 fm^{-3}$, the incompressibility has a value $K\simeq 15$ GeV whereas $\frac{v}{c} \simeq 0.84$.
From a study using 250 NS merger simulations, Perego et al. \cite{Perego2022} have indicated that an incompressibility $K\geq$12 GeV at the central density of the maximum mass NS in the binary
is favored to have a prompt collapse in BNSM to form BH, which is in line with the SEI model predictions.

The adiabatic index $\Gamma_{A}$ which is defined as,
\begin{eqnarray}
\Gamma_{A} = \left({\frac{\rho}{P}}{\frac{\partial P_{NSM}}{\partial {\rho}}}\right)_{\bar{S}},
\label{eq21b}
\end{eqnarray}
is shown in panel (c) of Fig.\ref{fig7} as a function of density for the three isoentropic EoSs of $\bar{S}$=1, 2, 3. The adiabatic index is a measure of the fractional variation of pressure and gives a measure of the stiffness of the EoS.
These three curves corresponding to the three considered EoSs are very similar to the MDI results, which are  shown in Fig. 6 of
Ref.\cite{Moustakidis2021}.
The central density of the maximum mass NSs for these three EoSs of constant lepton fraction increases as $\bar{S}$ decreases, as can be seen from Table \ref{tab2} which is also the finding for MDI in Ref.\cite{Moustakidis2021} as well as for the BBG calculations \cite{Burgio2010,Bombaci2021}.


 \subsection{Formation of neutron star in early stage}

\begin{figure}[t]
	\begin{center}
		\includegraphics[height=11cm,width=18cm]{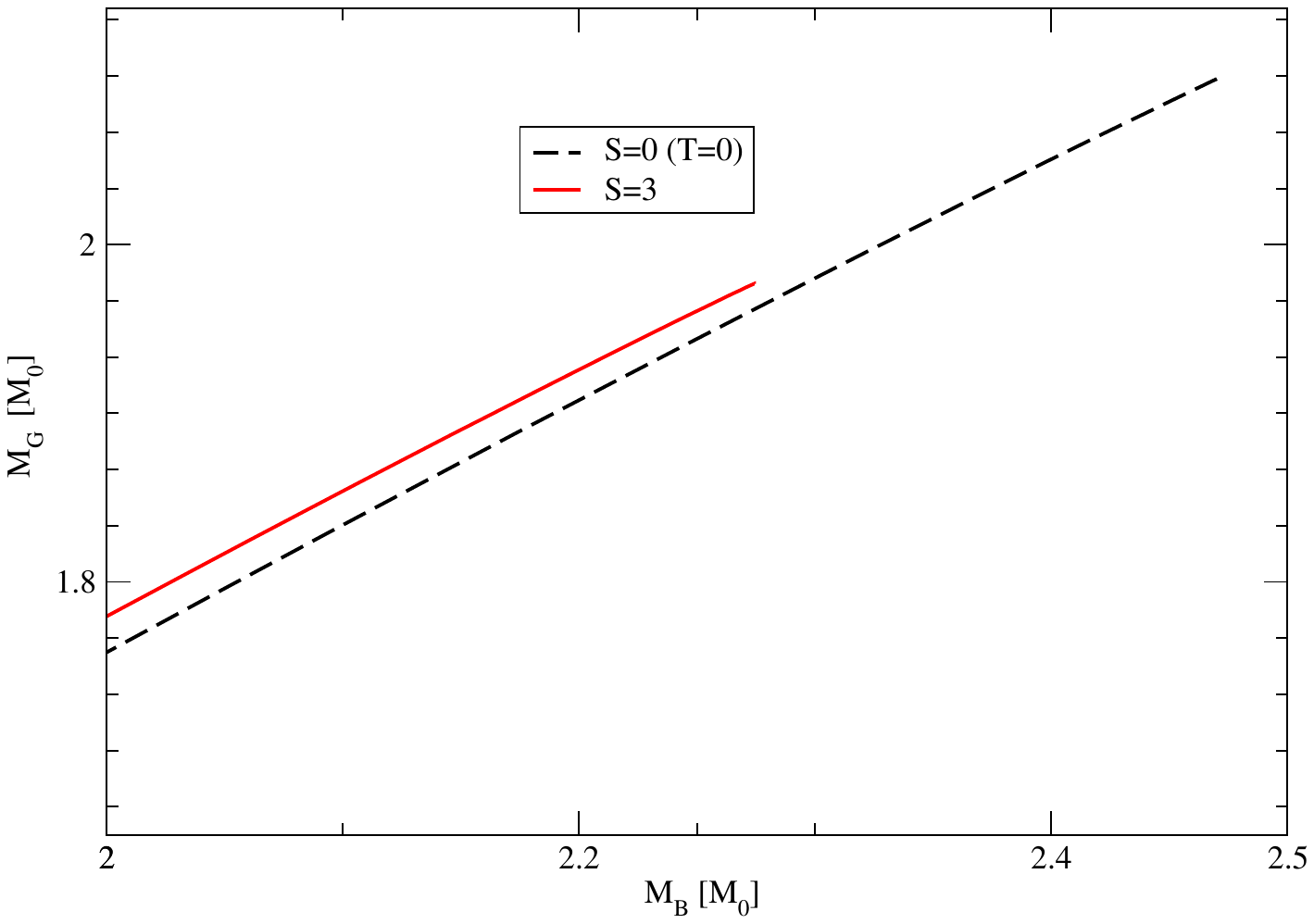}
		\caption{Proto-neutron star sequence for the isoentropic $S$=3 EoS and NS star mass sequence for the $T$=0 neutrino-free EoS of SEI ($\gamma$=2/3, $L$=75 MeV) as a function of baryonic mass $M_B$. }
		\label{fig11}
	\end{center}		
			 \end{figure}

The supernovae event where a PNS is formed, evolves into a NS and the process is covered in two steps. Mostly the whole mass accretion forming the PNS takes place within $t\sim$3 s of the core bounce at $t$=0. During this period of time the temperature of the PNS reaches few tens of MeV where the neutrinos are trapped. In this stage the EoS of the PNS is that of the hot isoentropic matter. In the subsequent $t\sim$ 30s (neutron diffusion time-scale) the neutrinos are diffused out and the star cools down to NS configuration, whose EoS is that of the $T$=0 neutrino-free NSM. During this transition no mass loss occurs and the baryonic mass (BM) remains almost the same. This is shown in Figure \ref{fig11} where we show the gravitational mass, $M_G$, of the proto-neutron star sequence for our isoentropic EoS of $\bar{S}$=3 and of the NS sequence of the $T$=0 neutrino-free EoS, both, as a function of the baryonic mass $M_B$. The end points of each curve depict the maximum mass for each sequence. We denote the maximum limits for GM and BM of PNS sequence as $M^{(i)}_{G{max}}$ and $M^{(i)}_{B{max}}$, and of NS sequence by $M^{(f)}_{G{max}}$ and $M^{(f)}_{B{max}}$. The $M_G$ and $M_B$ values in Table (\ref{tab2}) for the isoentropic EoSs correspond to the former, whereas, that of $T$=0, neutrino-free EoS corresponds to the later ones. A PNS having BM $M^{(i)}_{B} < M^{(i)}_{B{max}}$ will normally evolve to a NS of gravitational mass $M^{(f)}_{G}$ corresponding to the same baryonic mass $M^{(i)}_B$. The binding energy of the neutron star $\Delta{B}$=$(M^{(i)}_G-M^{(f)}_G)c^2$ is released mostly in the form of neutrinos. In the typical case of $M^{(i)}_B$=$M^{(i)}_{B{max}}$ of our Fig.\ref{fig11}, a NS of GM  $M^{(f)}_{G}=1.958M_{\odot}$ will be obtained with the release of $\Delta{B}=3.01{ \text{  }  } 10^{52}$ erg. If in a PNS the BM is $M^{(i)}_{B{max}} < M^{(i)}_B \le M^{(f)}_{B{max}}$ then it cannot be supported by the matter pressure to counterbalance the gravitational collapse, and it is likely that it will turn to a low-mass black hole after reaching supranuclear density. In the range $[M^{(i)}_{B{max}}, M^{(f)}_{B{max}}]$, there can be both heavier NSs and low-mass BHs. The heavier mass NSs can be formed by accretion of mass in a binary where there is a NS and the companion star is a normal star.

 \section{Summary and conclusion}

Our objective in the present work is to formulate the EoSs of hot $\beta$-stable neutron star n+p+e+$\mu$ matter in isothermal neutrino-free and isoentropic neutrino-trapped conditions. The study is done in the non-relativistic frame at mean field level using the finite range simple effective interaction with a Yukawa form-factor. At zero-temperature the EoS built up with this model conforms to the maximum mass and R$_{1.4}$ radius constraints. At finite temperature, the thermal effects in the mean field and the EoS are built upon the $T$=0 values through the FD distribution functions in the kinetic and exchange parts so that the laws of thermodynamics remain consistent in both zero- and finite temperature regimes.

The composition and the EoS of the neutrino-free isothermal hot $\beta$-stable NSM has been computed at both low and high temperatures relevant to PNS and BNSM events. The influence of temperature on the particle fractions and the EoS of NSM is found to be more prominent in the low density and higher temperature domain. At high temperature the thermal effects are gradually moderated with increase in density, approaching the zero-temperature limiting values. With increase in temperature, the charged particle fractions, Y$_p$, Y$_e$ and Y$_\mu$, increase while Y$_n$ decreases making the matter more symmetric. For a given temperature, the charge particle fractions, Y$_p$ and Y$_e$, have decreasing trend from very low density upto the muon threshold density beyond which these curves rise with increase in density. The muon production threshold density decreases with increase in $T$. The SEI results of particle fractions compare well with the BBG calculation results of Burgio and Schulze 2010 \cite{Burgio2010}. The free energy per particle in neutrino-free NSM at finite temperature decreases compared to the cold zero-temperature values making the EoS softer, in agreement with the results found in other model calculations. This is on account of the thermal energy at finite temperature, which is characterized by the entropy in the system. The entropy per particle in NSM at 5 and 50 MeV temperatures are shown in panel (b) of Fig.\ref{fig4} give an measure of thermal energy as temperature rises. In these curves the muon production threshold density is marked by a Breit-Wigner type resonance peak which shifts to lower density with increase in temperature. The thermal index $\Gamma_{th}$, is related to the ratio of the thermal pressure to the thermal internal energy of the constituent nucleons and leptons in NSM, is found to have almost a constant behaviour about $\Gamma_{th}\sim$ 1.5 irrespective of the temperature,
shown in Fig.\ref{fig6}. It justifies the $\Gamma$-law used in the BNSM simulation studies.

The studies on NSM and NSs found in the literature, both cold and hot, mostly use the parabolic approximation in ANM for solving the $\beta$-equilibrium condition to find the composition, as well as, for computing the various thermodynamical properties of the EoS. We have computed the equilibrium particle fractions and the thermodynamical quantities at finite temperature for neutrino-free NSM under the PA by performing calculations in SNM and PNM, and compared the results with the exact calculated data. It is found that the particle fraction composition as well as the thermodynamical properties of the EoS of NSM obtained under the PA compare well with the exactly evaluated results. This justifies that within the framework of SEI model, the PA is an accurate alternative to the exact calculation of hot NSM.

The formation of PNS in supernovae matter takes place where the matter is in neutrino-trapped hot isoentropic condition having entropy per particle in the range $\bar{S}$=1--3 (in the unit of $k_B$). With our SEI EoS the particle fractions and the EoSs of the $\nu$-trapped isoentropic matter evaluated for the entropy per particle values, $\bar{S}$=1, 2, 3, shown in Figs.\ref{fig8} and \ref{fig9}(a), exhibit almost the same results.
A similar situation is found in the results of the particle fractions predicted by the neutrino-trapped BBG calculation of Burgio and Schulze \cite{Burgio2011}. However, the
temperature profiles of the inside matter for these three different isoentropic EoSs show sharp difference. For higher values of $\bar{S}$ the temperature profiles of the isentropes are higher. The obvious reason is that higher entropy per particle implies higher disorderness in the system that can be attained by larger value of $T$. It is found that the isentropes have small dependence on the NM incompressibility, whereas, their dependence on the stiffness of the symmetry energy is quite substantial. As the slope parameter decreases the temperature profile of the isentropes increases sharply. In Fig.\ref{fig9}(b) this feature has been shown for the two EoSs having the slope parameters, $L$=75 and 65 MeV, where the rest of the NM saturation properties remain the same. This is because the baryonic number in a given mass NS resulting from the EoS having higher value of $L$ is higher, as a consequence, a lower $T$ is required to maintain the system at the same isoentropic condition compared to the one resulting from the lower $L$-EoS that contains less number of baryons.
The incompressibility, velocity of sound and adiabatic index for the three isoentropic EoSs of NSM have been computed where the influence of entropy on the former two are found to be very small. The velocity of sound remains causal for the three EoSs of $\bar{S}$=1, 2, 3 over the whole density range shown in panel (b) of Fig.\ref{fig7}. However, the adiabatic index shows marked entropy dependence which is similar to the results found for the MDI interaction in Ref.\cite{Moustakidis2021}. The maximum mass NSs predicted by the three isoentropic EoSs show an increasing trend where the central density decreases with increase in entropy per particle from 1 to 3, which is also the finding in other model calculations.
The early stage evolution of the PNS to NS has been discussed using our SEI EoS in the light of the work of Logoteta, Perego \& Bombaci (2021) \cite{Bombaci2021}and the possibility that a PNS can turn to a low-mass black hole is also pointed out. Our immediate interest is two fold, where we shall use the present EoS for the core-collapse supernovae simulation, and secondly we shall compute the bulk viscosity due to urca-processes in $\nu$-trapped NSM that can be used in the study of damping mechanism of the large oscillations produced in the BNSM.

\vspace{0.5cm}
\noindent
{\bf Acknowledgements}
\vspace{0.5cm}

    T.R.R. offers sincere thanks to Prof. B. Behera for useful discussions and acknowledges S. P. Pattnaik for assisting in computation. M. C. and X. V. acknowledge partial support from Grants No.
PID2020-118758GB-I00 and No. CEX2019-000918-M (through the ``Unit of
Excellence Mar\'{\i}a de Maeztu 2020-2023'' award to ICCUB) from the
Spanish MCIN/AEI/10.13039/501100011033.
DNB acknowledges support from SERB, DST, Government of India, through Grant No. CRG/2021/007333.

\end{document}